\documentclass[aps,prl,reprint,amsfonts,amsmath,amssymb,longbibliography]{revtex4-2}
\usepackage{xcolor}

\usepackage{hyperref}
\usepackage{graphicx}
\usepackage{bm}
\usepackage{newtxtext}
\usepackage[varg]{newtxmath}
\usepackage[normalem]{ulem}
\usepackage{lipsum}
\usepackage{slashed}
\DeclareMathOperator{\sh}{sh}
\DeclareMathOperator{\sech}{sech}
\DeclareMathOperator{\erf}{erf}

\hypersetup{colorlinks=true,linkcolor=blue,citecolor=blue,urlcolor=blue}

\begin{document}
\title{Theory for Fourier-limited attosecond pulse generation in solids}
\author{Shohei Imai}
\author{Atsushi Ono}
\affiliation{Department of Physics, Tohoku University, Sendai 980-8578, Japan}
\date{\today}

\begin{abstract}
The generation of ultrashort light pulses is essential for the advancement of attosecond science.
Here, we show that attosecond pulses approaching the Fourier limit can be generated through optimized optical driving of tunneling particles in solids.
We propose an ansatz for the wave function of tunneling electron--hole pairs based on a rigorous expression for massive Dirac fermions, which enables efficient optimization of the waveform of the driving field.
It is revealed that the dynamic sign change in the effective mass due to optical driving is crucial for shortening the pulse duration, which highlights a distinctive property of Bloch electrons that is not present in atomic gases, i.e., the periodic nature of crystals.
These results show the potential of utilizing solid materials as a source of attosecond pulses.
\end{abstract}
\maketitle

\textit{Introduction.}---%
Ultrafast physics has advanced significantly owing to the development of femtosecond lasers~\cite{Strickland2019}.
These intense, short pulses allow for time-resolved measurements of chemical reactions of molecules~\cite{Zewail1988} and even induce macroscopic phase transitions in condensed matter systems~\cite{Koshihara2022, Ishihara2019, Miyamoto2018b, Kirilyuk2010}.
In recent years, the pulse duration of lasers has reached attosecond time scales, enabling observation of coherent dynamics in atoms, molecules, and solids~\cite{Agostini2004, Krausz2009, Gallmann2012, Reid2016, Ramasesha2016, Li2020c, Pitruzzello2022}, and potentially lightwave control of quantum systems within a single optical cycle~\cite{Goulielmakis2007, Krausz2009, Krausz2014, Luu2015a, Reimann2018, Kawakami2018, Kawakami2018a, Schoetz2019a, Kawakami2020, Jimenez-Galan2021, Boolakee2022a}, which were previously inaccessible because of relaxation and dephasing and are now attracting considerable interest.

High harmonic generation (HHG) is a phenomenon in which a series of attosecond or femtosecond pulses with broad energy spectra is generated, and it has been actively studied owing to its fertile fundamental physics and potential applications in generating attosecond pulses.
HHG was first observed in atomic gases~\cite{McPherson1987, Ferray1988, Corkum1993, Lewenstein1994, Hentschel2001} and recently in solids such as semiconductors~\cite{Ghimire2011, Schubert2014a, Hohenleutner2015, Garg2016a, Langer2016, Langer2017, Huttner2017, Kruchinin2018, Ortmann2021, Xia2021, Tamaya2016}, topological materials~\cite{Yoshikawa2017, Bai2020a, Schmid2021, Lv2021}, and strongly correlated systems~\cite{Silva2018, Murakami2018d, Tancogne-Dejean2018a, Nag2019, Roy2020a, Murakami2021, Orthodoxou2021a, Udono2022a, Shao2022, Lysne2020a, Imai2019q, Zhu2021a, Fauseweh2020b, Takayoshi2019, Kanega2021a, Bionta2021, Uchida2021b, Murakami2022b, Granas2022, Alcala2022a}.
Condensed matter systems are promising pulse sources because their density is much higher than that of atomic gas systems, and research on attosecond pulse generation in such systems~\cite{Garg2016, Garg2018, Guan2016, Song2019, Nourbakhsh2021, Sadeghifaraz2022} has recently been triggered by the development of waveform synthesis of optical electric fields~\cite{Hassan2012, Mucke2015, Lin2020, Tian2021, Alqattan2022, Su2022}.

\begin{figure}[b]
\centering
\includegraphics[width=\columnwidth]{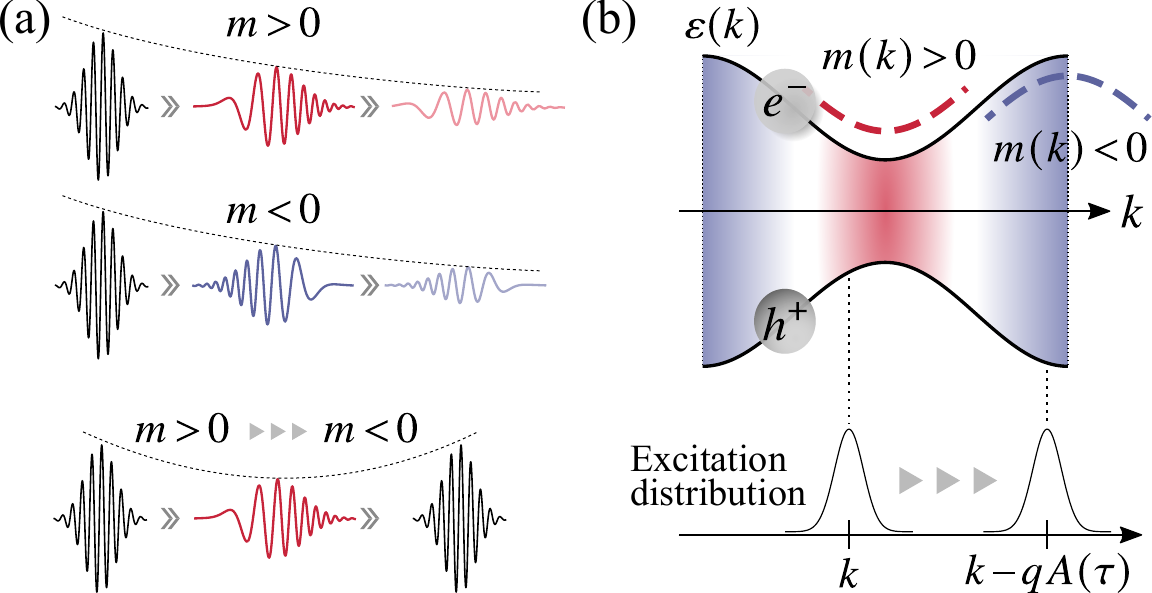}
\caption{(a)~Dynamics of wave packets in real space.
The wave packet of free or undriven particles monotonically spreads out (top two panels), whereas that of Bloch electrons can be refocused by optical driving (bottom panel).
(b)~Sketch of optically driven particles in crystalline solids.
The sign of the effective mass changes during optical driving.}
\label{fig:mass_sign}
\end{figure}

One of the difficulties in shortening the pulse duration lies in a chirp of wave packets.
Ionized or excited free electrons (holes) have a positive (negative) mass---i.e., chirp---and their sign does not change.
Therefore, the width of the wave packet monotonically spreads out as time evolves, which broadens the pulse width when carriers recombine, as depicted in Fig.~\ref{fig:mass_sign}(a).
In atomic gas systems, this issue can be resolved by passing pulses through a metal foil with negative dispersion, leading to a pulse duration of several tens of attoseconds~\cite{Gaumnitz2017, Shafir2012}.
On the other hand, in crystalline solids, the sign of the effective mass of carriers can be changed by optical driving [see Fig.~\ref{fig:mass_sign}(b)], suggesting an efficient way to refocus wave packets based on the inherent nature of Bloch electrons.

There are two contributions to the pulse emission from solids: one is intraband currents due to nonlinear changes in the carrier velocities, and the other is interband currents associated with recombination of the carrier wave packets.
Attosecond pulse generation based on intraband currents was reported in $\mathrm{SiO_2}$~\cite{Garg2016}, whereas that based on interband currents has received mainly theoretical attention~\cite{Guan2016, Song2019, Nourbakhsh2021} and is still not fully understood because, for instance, the theoretical description of quantum tunneling induced by intense electric fields is difficult~\cite{Vitanov1996, Oka2012a, Lewenstein1994, Shafir2012, MacColl1932, Yakaboylu2014, Kheifets2020}.
A recent theoretical study~\cite{Imai2022a} analyzed the pulse emission generated by recombination of resonantly excited carriers.
The results of that study imply that if carriers are excited through quantum tunneling and subsequently driven by an optical field, their energy spectrum will be broad, and the mass of the carriers can be reversed, which may lead to a short pulse width.

In this Letter, we investigate the real-time dynamics of tunneling electron--hole pairs in an intense electric-field pulse.
First, we make an ansatz for the wave function of the tunneling electron--hole pairs by expanding a rigorous formula for quantum tunneling of massive Dirac fermions~\cite{osti_4159359} into that of Bloch electrons in crystalline solids.
We then propose a method for optimizing the waveform of a driving pulse to shorten the emitted pulse duration.
A numerical simulation of the dynamics in a band insulator successfully demonstrates the generation of a short pulse near the Fourier limit, revealing that the dynamic sign change in the effective mass, which is unique to Bloch electrons, plays an essential role.

\textit{Ansatz for the wave function of tunneling particles.}---%
Given a conduction band and a valence band involved with quantum tunneling due to an intense pulse, we can consider a two-band effective model described by the Dirac equation,
\begin{align}
(\mathrm{i} \slashed{\partial} - mv) \psi = 0, \label{eq:dirac}
\end{align}
with $\hbar = 1$, where $\psi$ denotes the Dirac spinor for a fermion with mass $m$, and $v$ is the counterpart of the speed of light.
The positive-energy (negative-energy) solution of Eq.~\eqref{eq:dirac}, denoted by $\psi^{\mathrm{cb}}$ ($\psi^{\mathrm{vb}}$), corresponds to the conduction (valence) electron.
Narozhnyi and Nikishov~\cite{osti_4159359} derived rigorous expressions for $\psi^{\mathrm{cb}}$ and $\psi^{\mathrm{vb}}$ after tunneling of electrons induced by the Sauter potential, $A_{\mathrm{S}}(\tau) = - A_{\mathrm{S0}} [1 + \tanh(\omega_{\mathrm{S}} \tau)]/2$, where $\tau$ represents time.
The electric field of the Sauter potential, $E_{\mathrm{S}}(\tau) = - \partial_{\tau} A_{\mathrm{S}}(\tau) = A_{\mathrm{S0}} \omega_{\mathrm{S}} \sech^2(\omega_{\mathrm{S}} \tau)/2$, reaches its peak value of $A_{\mathrm{S}0} \omega_{\mathrm{S}}/2$ at $\tau = 0$ and induces electron tunneling.

In this work, we extend the Narozhnyi--Nikishov formula to encompass crystalline band insulators in a vector potential that is not restricted to the Sauter potential, by making an ansatz for the wave function of the tunneling electron--hole pair,
\begin{align}
& \psi^{\mathrm{vb}}_{k}(\tau)^* \psi^{\mathrm{cb}}_{k}(\tau) \notag \\
& \approx |\psi^{\mathrm{vb}}_k||\psi^{\mathrm{cb}}_k| \frac{ G(\mu+\nu-\lambda) G(-\mu+\nu+\lambda) G(-\mu+\nu-\lambda)}{G(-\mu-\nu-\lambda) G(2\nu)^2} \notag \\
&\quad \times \exp \biggl\{ \mathrm{i}\int_{0}^{\infty}\mathrm{d}\tau \left[ \varepsilon_\mathrm{D}(k-qA_\mathrm{S}(\tau)) - \varepsilon_\mathrm{D}(k-qA_\mathrm{S}(\infty)) \right] \biggr\} \notag \\
&\quad \times \exp \biggl[ -\mathrm{i}\int_{0}^{\tau}\mathrm{d}\tau' \varepsilon(k-qA(\tau')) \biggr]
\label{eq:bi_tunnel_cv}
\end{align}
for $\tau \gtrsim |\omega_{\mathrm{S}}^{-1}|$.
Here, $\varepsilon(k)$ represents the energy of the electron--hole pair with momentum $k$ in a band insulator, and $\varepsilon_{\mathrm{D}}(k) =2\sqrt{(mv^2)^2+(vk)^2}$ is that for Dirac fermions.
The absolute values of the wave functions are given by $|\psi_k^{\mathrm{cb}}| = \sqrt{\sh(\mu+\nu-\lambda) \sh(-\mu-\nu-\lambda) / \sh(-2\mu) \sh(2\nu)}$ and $|\psi_k^{\mathrm{vb}}| = \sqrt{\sh(-\mu+\nu+\lambda) \sh(-\mu+\nu-\lambda)/ \sh(-2\mu) \sh(2\nu)}$, where $\sh(x)$ is shorthand for $\sinh(\pi x)$.
The function $G$ is defined by $G(y) = \sqrt{y \sinh(\pi y)/\pi} \varGamma(\mathrm{i}y)$ for $y \in \mathbb{R}$ with $\varGamma$ being the gamma function, satisfying $|G| = 1$.
The parameters in Eq.~\eqref{eq:dirac} are given by $mv^2 = \varepsilon(k_0)/2$ and $v^2 = \varepsilon''(k_0) \varepsilon(k_0)/4$, and thereby $\varepsilon(k)$ can be approximated by the Dirac dispersion around each minimum of $\varepsilon(k)$ at $k = k_0$.
The other parameters, $\mu$, $\nu$, and $\lambda$, are determined by the relations, $2\omega_{\mathrm{S}}\mu = -\varepsilon_{\mathrm{D}}(k)/2$, $2\omega_{\mathrm{S}}\nu = \varepsilon_\mathrm{D}(k+qA_{\mathrm{S0}})/2$, and $2\omega_{\mathrm{S}} \lambda=-vqA_{\mathrm{S0}}$.
The electric charge, $q$, is set to one in our calculations.
Since we are interested in a situation where the electric field $E(\tau) = -\partial_\tau A(\tau)$ has a strong peak that causes electron tunneling, $A_{\mathrm{S0}}$ and $\omega_{\mathrm{S}}$ are set to $A_{\mathrm{S0}} = -2A(0)$ and $\omega_{\mathrm{S}} = -E(0)/A(0)$, with $\tau = 0$ being the time when $E(\tau)$ reaches its maximum.
Note that replacing $\varepsilon(k)$ with $\varepsilon_{\mathrm{D}}(k)$ and $A(\tau)$ with $A_{\mathrm{S}}(\tau)$ reproduces the result in Ref.~\cite{osti_4159359}.
To achieve attosecond pulse generation without unnecessary complications, we will assume that the tunneling pairs are driven within a Brillouin zone and experience tunneling only once so that Eq.~\eqref{eq:bi_tunnel_cv} can be applied independently to every tunneling pair around each minimum of $\varepsilon(k)$.

Our ansatz in Eq.~\eqref{eq:bi_tunnel_cv} can describe not only the asymptotic state of the electron--hole pair after quantum tunneling, but also the transient dynamics induced by optical driving, as shown later.
The phase factor in the second to last line of Eq.~\eqref{eq:bi_tunnel_cv} describes the dynamical phase attributed to the backward time evolution of Dirac fermions from time $\tau \to \infty$ to $\tau = 0$, and that in the last line is the dynamical phase of Bloch electrons from time $0$ to $\tau$; these are crucial for compensating for the discrepancy between the given dispersion relation $\varepsilon(k)$ and the Dirac dispersion relation $\varepsilon_{\mathrm{D}}(k)$.
Furthermore, even though Eq.~\eqref{eq:bi_tunnel_cv} is deduced from the analytical result for massive Dirac fermions, we can apply it to insulators with indirect band gaps since it involves only the difference between the energies of the conduction and valence bands, $\varepsilon(k)$, and also to two- or three-dimensional systems by considering one-dimensional carrier motions driven by a linearly polarized pulse~\cite{Imai2022a}.

\textit{Optimization of an optical driving field.}---%
We divide the vector potential of an external optical pulse, $A$, into two parts: an excitation pulse $A_{\mathrm{e}}$, which causes quantum tunneling of electrons, and a driving pulse $A_{\mathrm{d}}$, which drives the electron--hole pairs after tunneling.
The excitation pulse is assumed to have a strong peak at $\tau = 0$, while its waveform can be arbitrary.
The driving pulse reverses the relative group velocity of the electron--hole pair, resulting in pair recombination and pulse emission~\cite{Imai2022a}.
To minimize the duration of the emitted light pulse by optimizing the waveform of the driving pulse, we consider a set of the vector-potential values at several representative times, $\{A_{\mathrm{d}}(\tau_j)\}_{j=0,1,\dots, M}$ with $\tau_j = j \updelta\tau_{\mathrm{d}}$.
We reconstruct $A_{\mathrm{d}}(\tau)$ from $\{A_{\mathrm{d}}(\tau_j)\}$ by performing a fifth-order spline interpolation with boundary conditions: $A_{\mathrm{d}}(\tau_0) = \partial_\tau A_{\mathrm{d}}(\tau_0) = \partial_\tau^2 A_{\mathrm{d}}(\tau_0) = 0$, $A_{\mathrm{d}}(\tau_M) = -A_{\mathrm{e}}(\infty)$ [i.e., $A(\infty) = 0$], and $\partial_\tau A_{\mathrm{d}}(\tau_M) = \partial_\tau^2 A_{\mathrm{d}}(\tau_M) = 0$.

To take advantage of the analytical expression in Eq.~\eqref{eq:bi_tunnel_cv}, we rewrite Eq.~\eqref{eq:bi_tunnel_cv} as $\psi^{\mathrm{vb}}_{k}(\tau)^* \psi^{\mathrm{cb}}_{k}(\tau) = f_k \exp[\mathrm{i} \varphi_k(\tau)]$, where $f_k = |\psi^{\mathrm{vb}}_k||\psi^{\mathrm{cb}}_k|$ represents the amplitude, and the phase $\varphi_k(\tau)$ is a functional of $A_{\mathrm{d}}(\tau)$.
Since the expectation value of the interband electric current is given by a summation over $k$ of the product of $\psi^{\mathrm{vb}*}_{k} \psi^{\mathrm{cb}}_{k}$ and the transition dipole moment~\cite{Imai2022a}, the phase of the wave function (and thus that of the electric current), $\varphi_k$, should be $k$-independent when the electron and hole recombine at $\tau = \tau_{\mathrm{r}}$, for a minimum pulse duration.
We thus employ the Gauss--Newton method to minimize the residual sum of squares (RSS) defined by $R = N_{\mathrm{r}}^{-1} \sum_k \tilde{f}_k^n [\partial_k \varphi_k(\tau_\mathrm{r})]^2$.
Here, $N_{\mathrm{r}}$ represents the number of wavenumbers used for the optimization, $\tilde{f}_k \ (\propto f_k)$ denotes the normalized weight, and the exponent $n$ of $\tilde{f}_k$ is chosen such that the numerical procedure is kept stable.
In our calculations, we used $N_{\mathrm{r}} = 100$ and $n = 14$, and neglected eigenvalues smaller than $10^{-3}$ when solving the linear equations.
The gradient of RSS, $\updelta R/\updelta A_{\mathrm{d}}(\tau_j)$, is given by the product of $\updelta R/\updelta A_{\mathrm{d}}(\tau)$ and $\updelta A_{\mathrm{d}}(\tau)/\updelta A_{\mathrm{d}}(\tau_j)$; the latter derivative can be efficiently evaluated by replacing the spline function for $(A_{\mathrm{d}}(\tau_0), A_{\mathrm{d}}(\tau_1), \dots, A_{\mathrm{d}}(\tau_M))$ with that for $(0,\dots, 0, 1, 0, \dots, 0)$, where only the $j$th component is one.

\textit{Numerical results.}---%
To demonstrate the validity of the wave function in Eq.~\eqref{eq:bi_tunnel_cv} and the effectiveness of our optimization method, we consider a tight-binding model for a band insulator.
The Hamiltonian is given by $\mathcal{H} = \sum_{k} \vec{c}_k^\dagger H_k \vec{c}_k$, with
\begin{align}
H_k = -2 t_{\mathrm{h}} \cos(k)\, \sigma^x - (E_{\mathrm{g}}/2)\, \sigma^z,
\label{eq:bi_hamiltonian}
\end{align}
where $\vec{c}_k^\dagger = (c_{k1}^\dagger , c_{k2}^\dagger)$ is a vector of creation operators for electrons in orbital $\nu = 1, 2$ with crystal momentum $k$, and $\sigma^x$ and $\sigma^z$ denote the Pauli matrices.
The lattice constant $a$ is set to one.
The parameters $t_{\mathrm{h}}$ and $E_{\mathrm{g}}$ represent the transfer integral and the energy gap, respectively.
We used $t_{\mathrm{h}} = 3~\mathrm{eV}$ and $E_{\mathrm{g}} = 3~\mathrm{eV}$ as typical values for the band width ($\approx 10\ \mathrm{eV}$) and energy gap of semiconductors such as $\mathrm{ZnO}$.
The energy of the electron--hole pair is given by $\varepsilon(k) = 2\sqrt{(2t_{\mathrm{h}} \cos k )^2+(E_{\rm g}/2)^2}$, which has two minima at $k = \pm \pi/2\ (\equiv \pm k_0)$ in the first Brillouin zone, resulting in $mv^2 = E_{\mathrm{g}}/2$ and $v = +2t_{\mathrm{h}} \ (-2t_{\mathrm{h}})$ for $k=+k_0\ (-k_0)$.
In this model, it is sufficient to consider only one of the minima at $k = +k_0$; in general, we need to minimize the weighted sum of RSSs for all $k_0$'s.

The vector potential $A(\tau)$ is introduced through the Peierls substitution, $k \mapsto k-qA(\tau)$.
In particular, we chose an excitation pulse given by $A_{\mathrm{e}}(\tau) = -A_{\mathrm{e}0} [1 + \erf(\omega_{\mathrm{e}} \tau/\sqrt{2})]/2$, where $\erf$ is the error function, for which the Sauter-potential parameters are $A_{\mathrm{S}0} = A_{\mathrm{e}0}$ and $\omega_{\mathrm{S}} = \omega_{\mathrm{e}} \sqrt{2/\pi}$; we used $A_{\mathrm{e}0} = \pi/2$ and $\omega_{\mathrm{e}} = 0.2 t_{\mathrm{h}}$ so that the excitation distribution would be broad in energy.
The time evolution of the state $|\varPsi(\tau)\rangle$ is described by the equation $|\varPsi(\tau+\updelta\tau)\rangle = \exp[-\mathrm{i} \mathcal{H}(\tau + \updelta\tau/2) \updelta \tau] |\varPsi(\tau)\rangle + \mathcal{O}(\updelta \tau^3)$.
The electric current is defined by $J(\tau) = \langle \varPsi(\tau) \vert \hat{J}(\tau) \vert \varPsi(\tau) \rangle$ with $\hat{J}(\tau) = -N^{-1} \updelta \mathcal{H}(\tau)/\updelta A(\tau)$, where $N$ represents the number of $k$-points.
We employ $\tau_{\mathrm{r}} = 90 t_{\mathrm{h}}^{-1} \approx 20~\mathrm{fs}$, $\updelta \tau_{\mathrm{d}} = 15 t_{\mathrm{h}}^{-1}$, and $M = 6$ for the optimization, and $\updelta \tau = 0.03t_{\mathrm{h}}^{-1}$ and $N=1000$ for the simulation.
We set $t_{\mathrm{h}}/(qa) = 92~\mathrm{MV\, cm^{-1}}$ and $qt_{\mathrm{h}}/\hbar = 0.73~\mathrm{mA}$ as the units of the electric fields and electric currents, respectively, for $t_{\mathrm{h}} = 3~\mathrm{eV}$ and $a = 3.25~\mathrm{\mathring{A}}$.

\begin{figure}[t]
\centering
\includegraphics[width=\columnwidth]{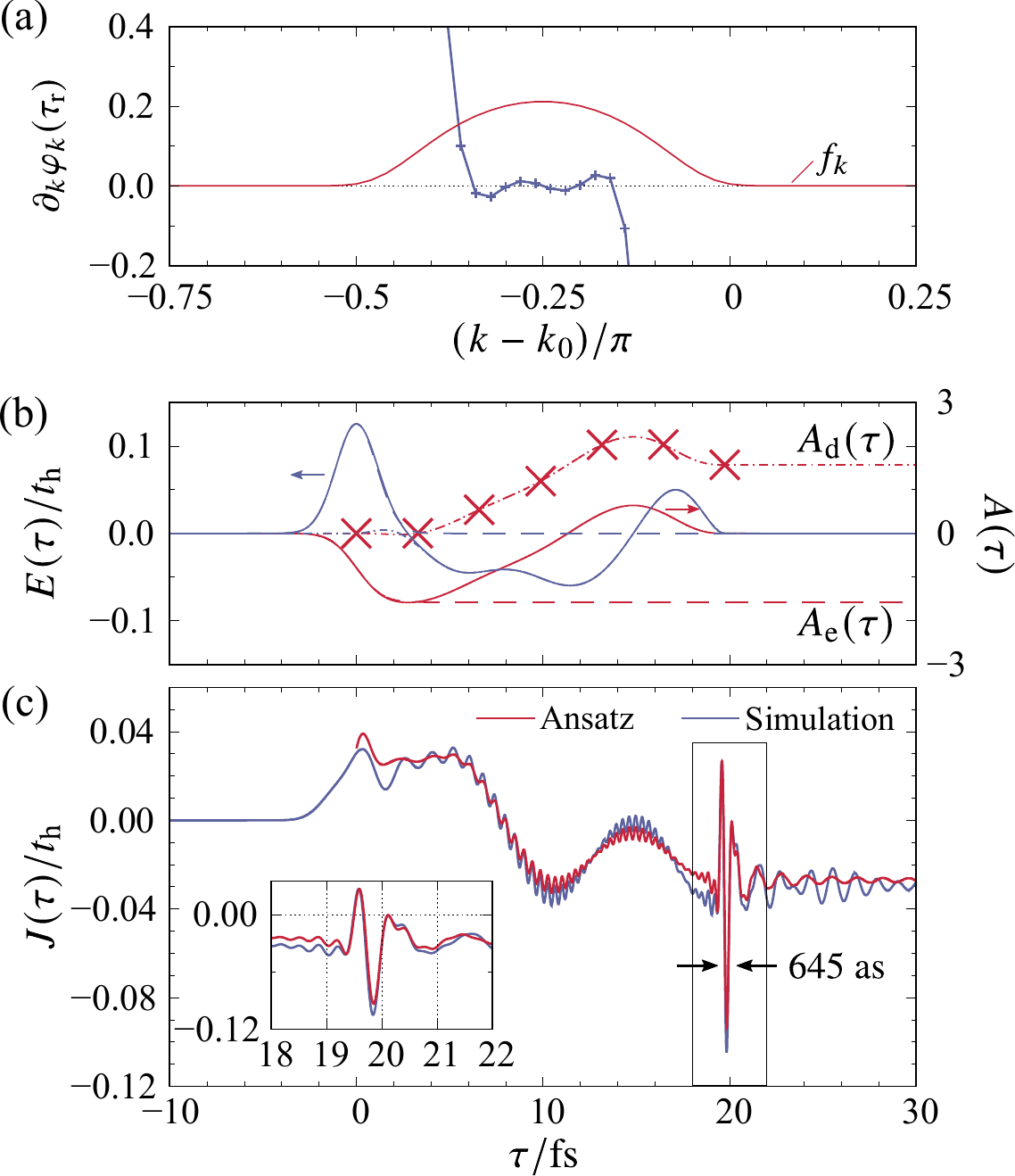}
\caption{(a)~Optimized phase derivative and excitation distribution at recombination time, $\tau_{\mathrm{r}} \approx 20~\mathrm{fs}$.
(b)~Optimized waveforms of the electric field (blue curve) and vector potential (red curves).
The crosses indicate the representatives $\{A_{\mathrm{d}}(\tau_j)\}$ used for optimization.
(c)~Electric currents calculated from the ansatz (red) and from the numerical simulation (blue).
The inset is an enlarged view from $\tau = 18~\mathrm{fs}$ to $22~\mathrm{fs}$.}
\label{fig:electric_current}
\end{figure}

Figure~\ref{fig:electric_current}(a) shows the excitation distribution of the electron--hole pair, $f_k$, and the optimized phase derivative, $\partial_k \varphi_k$, at time $\tau = \tau_{\mathrm{r}}$.
We found that the phase derivative was optimized, i.e., $\partial_k \varphi_k \approx 0$, around the center of the excitation distribution, $k-k_0 = k_{\mathrm{c}} = -qA_{\mathrm{S}0}/2 = -\pi/4$, with $R = 2.18 \times 10^{-5}$.
The optimized waveforms of the electric field and vector potential are plotted in Fig.~\ref{fig:electric_current}(b), where we took seven representatives $\{ A_{\mathrm{d}}(\tau_{j}) \}_{j=0,1,\dots,6}$ as indicated by the crosses.
Figure~\ref{fig:electric_current}(c) displays the electric current $J(\tau)$ calculated from our ansatz in Eq.~\eqref{eq:bi_tunnel_cv} and that from the numerical simulation of the tight-binding model.
The electric field reaches its maximum at $\tau = 0$, and the electric current associated with the tunneling particles flows simultaneously.
Following the optical driving, a sub-cycle electric current pulse is generated at $\tau \approx 20~\mathrm{fs}$ as a result of the recombination of the electron--hole pair.
The full width at half maximum (FWHM) of the pulse is evaluated to be $645~\mathrm{as}$.
There is good agreement between the ansatz and the numerical simulation, which indicates that the wave function in Eq.~\eqref{eq:bi_tunnel_cv} can accurately describe not only the asymptotic state in the long-time limit but also the transient state under optical driving.

\begin{figure}[t]
\centering
\includegraphics[width=\columnwidth]{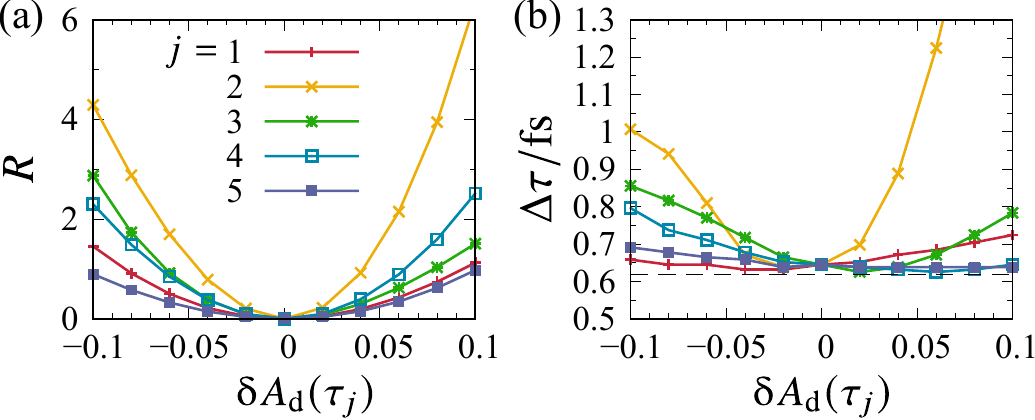}
\caption{(a)~RSS as a function of $\updelta A_{\mathrm{d}}(\tau_j)$.
(b)~FWHM of the generated pulse.
The dashed line indicates the Fourier-limited pulse duration.}
\label{fig:rss_delta_tau}
\end{figure}

Figure~\ref{fig:rss_delta_tau} shows the effects of an error in the representative $A_{\mathrm{d}}(\tau_j)$ on the RSS, $R$, and the FWHM of the generated pulse calculated from the ansatz, $\Delta \tau$, when one of the representatives deviates from the optimized value $A_{\mathrm{d}}(\tau_j)$ to $A_{\mathrm{d}}(\tau_j) + \updelta A_{\mathrm{d}}(\tau_j)$.
The RSS grows quadratically with $\updelta A_{\mathrm{d}}(\tau_j)$, indicating that the optimization procedure was performed successfully.
Furthermore, we find in Fig.~\ref{fig:rss_delta_tau}(b) that the FWHM of the optimized pulse is close to the Fourier limit ($\approx 619~\mathrm{as}$), which is evaluated by replacing $\psi_{k}^{\mathrm{vb}*} \psi_{k}^{\mathrm{cb}}$ with $f_k \exp[-\mathrm{i} \varepsilon(k) (\tau-\tau_{\mathrm{r}})]$, and is rather insensitive to $\updelta A_{\mathrm{d}}(\tau_j)$~\footnote{The minimum FWHM ($\approx 625~\mathrm{as}$) is located away from $\updelta A_{\mathrm{d}} = 0$, because we used the same $\tau_{\mathrm{r}}$ for $|\updelta A_{\mathrm{d}}| \neq 0$.}.
If one needs to generate a pulse with a FWHM less than $1~\mathrm{fs}$, the allowable error in the vector potential is estimated to be $|\updelta A(\tau_j) / A(\tau_j)| \lesssim 0.03$ for $\tau = \tau_{j = 2}$.

\textit{Discussion.}---%
To gain insight into the underlying mechanism of the short-pulse generation demonstrated above, let us consider a case where only two parameters---e.g., $\tau_{\mathrm{r}}$ and $A_{\mathrm{d}}(\tau_1)$ when $M=2$---are optimized.
In this case, the condition that $R$ is minimized can be written as
\begin{align}
\partial_k \varphi_k(\tau_\mathrm{r}) |_{k=k_\mathrm{c}} &= 0, \label{eq:1st_derivative} \\
\partial_k^2 \varphi_k(\tau_\mathrm{r}) |_{k=k_\mathrm{c}} &= 0, \label{eq:2nd_derivative}
\end{align}
where we set $k_0 = 0$ for simplicity.
Since $\varphi_k(\tau_{\mathrm{r}})$ can be expressed as $\varphi_k(\tau) = \theta_k - \int_{0}^{\tau} \mathrm{d}\tau'\, \varepsilon(k-qA(\tau'))$ with $\theta_k$ being a time-independent constant, Eq.~\eqref{eq:1st_derivative} reduces to $-\partial_k \theta_k \vert_{k=k_{\mathrm{c}}} + \int_0^{\tau_{\mathrm{r}}} \mathrm{d}\tau\, \partial_k \varepsilon(k-qA(\tau)) = 0$, which means that the relative displacement between the electron and hole is zero at $\tau = \tau_{\mathrm{r}}$, given that $-\partial_k \theta_k \vert_{k=k_{\mathrm{c}}}$ corresponds to the relative displacement at $\tau = 0$ as discussed in Refs.~\cite{Lewenstein1994, Geissler1999, Yakaboylu2014, Jurgens2020, Wang2021f}.

Equation~\eqref{eq:2nd_derivative} states that the chirp of the wave packet of the electron--hole pair is zero at $\tau = \tau_{\mathrm{r}}$, and it can be written as $\int_{0}^{\tau_{\mathrm{r}}} \mathrm{d}\tau\, m^{-1}(k-qA(\tau)) = \partial_k^2 \theta_k \vert_{k=k_{\mathrm{c}}}$, where $m^{-1}$ denotes the inverse mass defined by $m^{-1}(k) = \partial_k^2 \varepsilon(k)$.
The right-hand side is a constant with an approximate value of $0.57$ in the present calculation.
On the other hand, for massive Dirac fermions with $m_{\mathrm{D}}^{-1}(k) = \partial_k^2 \varepsilon_{\mathrm{D}}(k) > 0$, the time integral of the inverse mass monotonically increases and is estimated to be of order $10^2$; therefore, the condition in Eq.~\eqref{eq:2nd_derivative} is not satisfied.
As a result, the wave packet of free particles gradually spreads out in real space [the top two panels of Fig.~\ref{fig:mass_sign}(a)].
In crystalline solids, however, the sign of the effective mass can change during optical driving, as depicted in Fig.~\ref{fig:mass_sign}(b), which enables a reduction in the pulse duration by refocusing the wave packet [the bottom panel of Fig.~\ref{fig:mass_sign}(a)].

Finally, let us discuss the effect of dephasing on the short-pulse generation.
Since the fastest time scale of dephasing or relaxation in solids is typically on the order of ten femtoseconds, the coherence of electrons must be maintained on that time scale to observe intriguing ultrafast phenomena.
If we consider a dephasing time $T_2$ that does not depend on the wavenumbers of the carriers, the interband current will decay exponentially as $J(\tau) \propto \exp(-\tau/T_2)$~\cite{Vampa2014}, while its waveform---i.e., pulse duration---remains unchanged.
However, a more profound study that explicitly considers dephasing will be left for future work.

\textit{Conclusion.}---%
We demonstrated that attosecond pulses can be generated by controlling the motion of tunneling particles in solids by using optical driving.
The ansatz for the wave function of the tunneling particles in Eq.~\eqref{eq:bi_tunnel_cv} facilitates efficient optimization of the optical driving pulse, which results in a substantial reduction in the pulse duration towards the Fourier limit.
The present results provide a theoretical limit for the pulse duration and suggest that crystalline solids are promising sources of attosecond light pulses in combination with waveform synthesis of optical electric fields~\cite{Hassan2012, Mucke2015, Lin2020, Tian2021, Alqattan2022, Su2022}.

\begin{acknowledgments}
This work was supported by JSPS KAKENHI, Grant No.~JP21J10575, No.~JP18H05208, and No.~JP20K14394.
Some of the numerical calculations were performed using the facilities of the Supercomputer Center, the Institute for Solid State Physics, The University of Tokyo.
\end{acknowledgments}

\bibliography{reference}

\begin{thebibliography}{89}%
\makeatletter
\providecommand \@ifxundefined [1]{%
 \@ifx{#1\undefined}
}%
\providecommand \@ifnum [1]{%
 \ifnum #1\expandafter \@firstoftwo
 \else \expandafter \@secondoftwo
 \fi
}%
\providecommand \@ifx [1]{%
 \ifx #1\expandafter \@firstoftwo
 \else \expandafter \@secondoftwo
 \fi
}%
\providecommand \natexlab [1]{#1}%
\providecommand \enquote  [1]{``#1''}%
\providecommand \bibnamefont  [1]{#1}%
\providecommand \bibfnamefont [1]{#1}%
\providecommand \citenamefont [1]{#1}%
\providecommand \href@noop [0]{\@secondoftwo}%
\providecommand \href [0]{\begingroup \@sanitize@url \@href}%
\providecommand \@href[1]{\@@startlink{#1}\@@href}%
\providecommand \@@href[1]{\endgroup#1\@@endlink}%
\providecommand \@sanitize@url [0]{\catcode `\\12\catcode `\$12\catcode
  `\&12\catcode `\#12\catcode `\^12\catcode `\_12\catcode `\%12\relax}%
\providecommand \@@startlink[1]{}%
\providecommand \@@endlink[0]{}%
\providecommand \url  [0]{\begingroup\@sanitize@url \@url }%
\providecommand \@url [1]{\endgroup\@href {#1}{\urlprefix }}%
\providecommand \urlprefix  [0]{URL }%
\providecommand \Eprint [0]{\href }%
\providecommand \doibase [0]{https://doi.org/}%
\providecommand \selectlanguage [0]{\@gobble}%
\providecommand \bibinfo  [0]{\@secondoftwo}%
\providecommand \bibfield  [0]{\@secondoftwo}%
\providecommand \translation [1]{[#1]}%
\providecommand \BibitemOpen [0]{}%
\providecommand \bibitemStop [0]{}%
\providecommand \bibitemNoStop [0]{.\EOS\space}%
\providecommand \EOS [0]{\spacefactor3000\relax}%
\providecommand \BibitemShut  [1]{\csname bibitem#1\endcsname}%
\let\auto@bib@innerbib\@empty
\bibitem [{\citenamefont {Strickland}(2019)}]{Strickland2019}%
  \BibitemOpen
  \bibfield  {author} {\bibinfo {author} {\bibfnamefont {D.}~\bibnamefont
  {Strickland}},\ }\bibfield  {title} {\bibinfo {title} {{Nobel Lecture:
  Generating high-intensity ultrashort optical pulses}},\ }\href
  {https://doi.org/10.1103/RevModPhys.91.030502} {\bibfield  {journal}
  {\bibinfo  {journal} {Rev. Mod. Phys.}\ }\textbf {\bibinfo {volume} {91}},\
  \bibinfo {pages} {030502} (\bibinfo {year} {2019})}\BibitemShut {NoStop}%
\bibitem [{\citenamefont {Zewail}(1988)}]{Zewail1988}%
  \BibitemOpen
  \bibfield  {author} {\bibinfo {author} {\bibfnamefont {A.~H.}\ \bibnamefont
  {Zewail}},\ }\bibfield  {title} {\bibinfo {title} {{Laser Femtochemistry}},\
  }\href {https://doi.org/10.1126/science.242.4886.1645} {\bibfield  {journal}
  {\bibinfo  {journal} {Science}\ }\textbf {\bibinfo {volume} {242}},\ \bibinfo
  {pages} {1645} (\bibinfo {year} {1988})}\BibitemShut {NoStop}%
\bibitem [{\citenamefont {Koshihara}\ \emph {et~al.}(2022)\citenamefont
  {Koshihara}, \citenamefont {Ishikawa}, \citenamefont {Okimoto}, \citenamefont
  {Onda}, \citenamefont {Fukaya}, \citenamefont {Hada}, \citenamefont
  {Hayashi}, \citenamefont {Ishihara},\ and\ \citenamefont
  {Luty}}]{Koshihara2022}%
  \BibitemOpen
  \bibfield  {author} {\bibinfo {author} {\bibfnamefont {S.}~\bibnamefont
  {Koshihara}}, \bibinfo {author} {\bibfnamefont {T.}~\bibnamefont {Ishikawa}},
  \bibinfo {author} {\bibfnamefont {Y.}~\bibnamefont {Okimoto}}, \bibinfo
  {author} {\bibfnamefont {K.}~\bibnamefont {Onda}}, \bibinfo {author}
  {\bibfnamefont {R.}~\bibnamefont {Fukaya}}, \bibinfo {author} {\bibfnamefont
  {M.}~\bibnamefont {Hada}}, \bibinfo {author} {\bibfnamefont {Y.}~\bibnamefont
  {Hayashi}}, \bibinfo {author} {\bibfnamefont {S.}~\bibnamefont {Ishihara}},\
  and\ \bibinfo {author} {\bibfnamefont {T.}~\bibnamefont {Luty}},\ }\bibfield
  {title} {\bibinfo {title} {{Challenges for developing photo-induced phase
  transition (PIPT) systems: From classical (incoherent) to quantum (coherent)
  control of PIPT dynamics}},\ }\href
  {https://doi.org/10.1016/j.physrep.2021.10.003} {\bibfield  {journal}
  {\bibinfo  {journal} {Phys. Rep.}\ }\textbf {\bibinfo {volume} {942}},\
  \bibinfo {pages} {1} (\bibinfo {year} {2022})}\BibitemShut {NoStop}%
\bibitem [{\citenamefont {Ishihara}(2019)}]{Ishihara2019}%
  \BibitemOpen
  \bibfield  {author} {\bibinfo {author} {\bibfnamefont {S.}~\bibnamefont
  {Ishihara}},\ }\bibfield  {title} {\bibinfo {title} {{Photoinduced Ultrafast
  Phenomena in Correlated Electron Magnets}},\ }\href
  {https://doi.org/10.7566/JPSJ.88.072001} {\bibfield  {journal} {\bibinfo
  {journal} {J. Phys. Soc. Jpn.}\ }\textbf {\bibinfo {volume} {88}},\ \bibinfo
  {pages} {072001} (\bibinfo {year} {2019})}\BibitemShut {NoStop}%
\bibitem [{\citenamefont {Miyamoto}\ \emph {et~al.}(2018)\citenamefont
  {Miyamoto}, \citenamefont {Yamakawa}, \citenamefont {Morimoto},\ and\
  \citenamefont {Okamoto}}]{Miyamoto2018b}%
  \BibitemOpen
  \bibfield  {author} {\bibinfo {author} {\bibfnamefont {T.}~\bibnamefont
  {Miyamoto}}, \bibinfo {author} {\bibfnamefont {H.}~\bibnamefont {Yamakawa}},
  \bibinfo {author} {\bibfnamefont {T.}~\bibnamefont {Morimoto}},\ and\
  \bibinfo {author} {\bibfnamefont {H.}~\bibnamefont {Okamoto}},\ }\bibfield
  {title} {\bibinfo {title} {{Control of electronic states by a nearly
  monocyclic terahertz electric-field pulse in organic correlated electron
  materials}},\ }\href {https://doi.org/10.1088/1361-6455/aad023} {\bibfield
  {journal} {\bibinfo  {journal} {J. Phys. B}\ }\textbf {\bibinfo {volume}
  {51}},\ \bibinfo {pages} {162001} (\bibinfo {year} {2018})}\BibitemShut
  {NoStop}%
\bibitem [{\citenamefont {Kirilyuk}\ \emph {et~al.}(2010)\citenamefont
  {Kirilyuk}, \citenamefont {Kimel},\ and\ \citenamefont
  {Rasing}}]{Kirilyuk2010}%
  \BibitemOpen
  \bibfield  {author} {\bibinfo {author} {\bibfnamefont {A.}~\bibnamefont
  {Kirilyuk}}, \bibinfo {author} {\bibfnamefont {A.~V.}\ \bibnamefont
  {Kimel}},\ and\ \bibinfo {author} {\bibfnamefont {T.}~\bibnamefont
  {Rasing}},\ }\bibfield  {title} {\bibinfo {title} {{Ultrafast optical
  manipulation of magnetic order}},\ }\href
  {https://doi.org/10.1103/RevModPhys.82.2731} {\bibfield  {journal} {\bibinfo
  {journal} {Rev. Mod. Phys.}\ }\textbf {\bibinfo {volume} {82}},\ \bibinfo
  {pages} {2731} (\bibinfo {year} {2010})}\BibitemShut {NoStop}%
\bibitem [{\citenamefont {Agostini}\ and\ \citenamefont
  {DiMauro}(2004)}]{Agostini2004}%
  \BibitemOpen
  \bibfield  {author} {\bibinfo {author} {\bibfnamefont {P.}~\bibnamefont
  {Agostini}}\ and\ \bibinfo {author} {\bibfnamefont {L.~F.}\ \bibnamefont
  {DiMauro}},\ }\bibfield  {title} {\bibinfo {title} {{The physics of
  attosecond light pulses}},\ }\href
  {https://doi.org/10.1088/0034-4885/67/6/R01} {\bibfield  {journal} {\bibinfo
  {journal} {Rep. Prog. Phys.}\ }\textbf {\bibinfo {volume} {67}},\ \bibinfo
  {pages} {813} (\bibinfo {year} {2004})}\BibitemShut {NoStop}%
\bibitem [{\citenamefont {Krausz}\ and\ \citenamefont
  {Ivanov}(2009)}]{Krausz2009}%
  \BibitemOpen
  \bibfield  {author} {\bibinfo {author} {\bibfnamefont {F.}~\bibnamefont
  {Krausz}}\ and\ \bibinfo {author} {\bibfnamefont {M.}~\bibnamefont
  {Ivanov}},\ }\bibfield  {title} {\bibinfo {title} {{Attosecond physics}},\
  }\href {https://doi.org/10.1103/RevModPhys.81.163} {\bibfield  {journal}
  {\bibinfo  {journal} {Rev. Mod. Phys.}\ }\textbf {\bibinfo {volume} {81}},\
  \bibinfo {pages} {163} (\bibinfo {year} {2009})}\BibitemShut {NoStop}%
\bibitem [{\citenamefont {Gallmann}\ \emph {et~al.}(2012)\citenamefont
  {Gallmann}, \citenamefont {Cirelli},\ and\ \citenamefont
  {Keller}}]{Gallmann2012}%
  \BibitemOpen
  \bibfield  {author} {\bibinfo {author} {\bibfnamefont {L.}~\bibnamefont
  {Gallmann}}, \bibinfo {author} {\bibfnamefont {C.}~\bibnamefont {Cirelli}},\
  and\ \bibinfo {author} {\bibfnamefont {U.}~\bibnamefont {Keller}},\
  }\bibfield  {title} {\bibinfo {title} {{Attosecond Science: Recent Highlights
  and Future Trends}},\ }\href
  {https://doi.org/10.1146/annurev-physchem-032511-143702} {\bibfield
  {journal} {\bibinfo  {journal} {Annu. Rev. Phys. Chem.}\ }\textbf {\bibinfo
  {volume} {63}},\ \bibinfo {pages} {447} (\bibinfo {year} {2012})}\BibitemShut
  {NoStop}%
\bibitem [{\citenamefont {Reid}\ \emph {et~al.}(2016)\citenamefont {Reid},
  \citenamefont {Heyl}, \citenamefont {Thomson}, \citenamefont {Trebino},
  \citenamefont {Steinmeyer}, \citenamefont {Fielding}, \citenamefont
  {Holzwarth}, \citenamefont {Zhang}, \citenamefont {Del'Haye}, \citenamefont
  {S{\"{u}}dmeyer}, \citenamefont {Mourou}, \citenamefont {Tajima},
  \citenamefont {Faccio}, \citenamefont {Harren},\ and\ \citenamefont
  {Cerullo}}]{Reid2016}%
  \BibitemOpen
  \bibfield  {author} {\bibinfo {author} {\bibfnamefont {D.~T.}\ \bibnamefont
  {Reid}}, \bibinfo {author} {\bibfnamefont {C.~M.}\ \bibnamefont {Heyl}},
  \bibinfo {author} {\bibfnamefont {R.~R.}\ \bibnamefont {Thomson}}, \bibinfo
  {author} {\bibfnamefont {R.}~\bibnamefont {Trebino}}, \bibinfo {author}
  {\bibfnamefont {G.}~\bibnamefont {Steinmeyer}}, \bibinfo {author}
  {\bibfnamefont {H.~H.}\ \bibnamefont {Fielding}}, \bibinfo {author}
  {\bibfnamefont {R.}~\bibnamefont {Holzwarth}}, \bibinfo {author}
  {\bibfnamefont {Z.}~\bibnamefont {Zhang}}, \bibinfo {author} {\bibfnamefont
  {P.}~\bibnamefont {Del'Haye}}, \bibinfo {author} {\bibfnamefont
  {T.}~\bibnamefont {S{\"{u}}dmeyer}}, \bibinfo {author} {\bibfnamefont
  {G.}~\bibnamefont {Mourou}}, \bibinfo {author} {\bibfnamefont
  {T.}~\bibnamefont {Tajima}}, \bibinfo {author} {\bibfnamefont
  {D.}~\bibnamefont {Faccio}}, \bibinfo {author} {\bibfnamefont {F.~J.~M.}\
  \bibnamefont {Harren}},\ and\ \bibinfo {author} {\bibfnamefont
  {G.}~\bibnamefont {Cerullo}},\ }\bibfield  {title} {\bibinfo {title}
  {{Roadmap on ultrafast optics}},\ }\href
  {https://doi.org/10.1088/2040-8978/18/9/093006} {\bibfield  {journal}
  {\bibinfo  {journal} {J. Opt.}\ }\textbf {\bibinfo {volume} {18}},\ \bibinfo
  {pages} {093006} (\bibinfo {year} {2016})}\BibitemShut {NoStop}%
\bibitem [{\citenamefont {Ramasesha}\ \emph {et~al.}(2016)\citenamefont
  {Ramasesha}, \citenamefont {Leone},\ and\ \citenamefont
  {Neumark}}]{Ramasesha2016}%
  \BibitemOpen
  \bibfield  {author} {\bibinfo {author} {\bibfnamefont {K.}~\bibnamefont
  {Ramasesha}}, \bibinfo {author} {\bibfnamefont {S.~R.}\ \bibnamefont
  {Leone}},\ and\ \bibinfo {author} {\bibfnamefont {D.~M.}\ \bibnamefont
  {Neumark}},\ }\bibfield  {title} {\bibinfo {title} {{Real-Time Probing of
  Electron Dynamics Using Attosecond Time-Resolved Spectroscopy}},\ }\href
  {https://doi.org/10.1146/annurev-physchem-040215-112025} {\bibfield
  {journal} {\bibinfo  {journal} {Annu. Rev. Phys. Chem.}\ }\textbf {\bibinfo
  {volume} {67}},\ \bibinfo {pages} {41} (\bibinfo {year} {2016})}\BibitemShut
  {NoStop}%
\bibitem [{\citenamefont {Li}\ \emph {et~al.}(2020)\citenamefont {Li},
  \citenamefont {Lu}, \citenamefont {Chew}, \citenamefont {Han}, \citenamefont
  {Li}, \citenamefont {Wu}, \citenamefont {Wang}, \citenamefont {Ghimire},\
  and\ \citenamefont {Chang}}]{Li2020c}%
  \BibitemOpen
  \bibfield  {author} {\bibinfo {author} {\bibfnamefont {J.}~\bibnamefont
  {Li}}, \bibinfo {author} {\bibfnamefont {J.}~\bibnamefont {Lu}}, \bibinfo
  {author} {\bibfnamefont {A.}~\bibnamefont {Chew}}, \bibinfo {author}
  {\bibfnamefont {S.}~\bibnamefont {Han}}, \bibinfo {author} {\bibfnamefont
  {J.}~\bibnamefont {Li}}, \bibinfo {author} {\bibfnamefont {Y.}~\bibnamefont
  {Wu}}, \bibinfo {author} {\bibfnamefont {H.}~\bibnamefont {Wang}}, \bibinfo
  {author} {\bibfnamefont {S.}~\bibnamefont {Ghimire}},\ and\ \bibinfo {author}
  {\bibfnamefont {Z.}~\bibnamefont {Chang}},\ }\bibfield  {title} {\bibinfo
  {title} {{Attosecond science based on high harmonic generation from gases and
  solids}},\ }\href {https://doi.org/10.1038/s41467-020-16480-6} {\bibfield
  {journal} {\bibinfo  {journal} {Nat. Commun.}\ }\textbf {\bibinfo {volume}
  {11}},\ \bibinfo {pages} {2748} (\bibinfo {year} {2020})}\BibitemShut
  {NoStop}%
\bibitem [{\citenamefont {Pitruzzello}(2022)}]{Pitruzzello2022}%
  \BibitemOpen
  \bibfield  {author} {\bibinfo {author} {\bibfnamefont {G.}~\bibnamefont
  {Pitruzzello}},\ }\bibfield  {title} {\bibinfo {title} {{A bright future for
  attosecond physics}},\ }\href {https://doi.org/10.1038/s41566-022-01031-w}
  {\bibfield  {journal} {\bibinfo  {journal} {Nat. Photonics}\ }\textbf
  {\bibinfo {volume} {16}},\ \bibinfo {pages} {550} (\bibinfo {year}
  {2022})}\BibitemShut {NoStop}%
\bibitem [{\citenamefont {Goulielmakis}\ \emph {et~al.}(2007)\citenamefont
  {Goulielmakis}, \citenamefont {Yakovlev}, \citenamefont {Cavalieri},
  \citenamefont {Uiberacker}, \citenamefont {Pervak}, \citenamefont
  {Apolonski}, \citenamefont {Kienberger}, \citenamefont {Kleineberg},\ and\
  \citenamefont {Krausz}}]{Goulielmakis2007}%
  \BibitemOpen
  \bibfield  {author} {\bibinfo {author} {\bibfnamefont {E.}~\bibnamefont
  {Goulielmakis}}, \bibinfo {author} {\bibfnamefont {V.~S.}\ \bibnamefont
  {Yakovlev}}, \bibinfo {author} {\bibfnamefont {A.~L.}\ \bibnamefont
  {Cavalieri}}, \bibinfo {author} {\bibfnamefont {M.}~\bibnamefont
  {Uiberacker}}, \bibinfo {author} {\bibfnamefont {V.}~\bibnamefont {Pervak}},
  \bibinfo {author} {\bibfnamefont {A.}~\bibnamefont {Apolonski}}, \bibinfo
  {author} {\bibfnamefont {R.}~\bibnamefont {Kienberger}}, \bibinfo {author}
  {\bibfnamefont {U.}~\bibnamefont {Kleineberg}},\ and\ \bibinfo {author}
  {\bibfnamefont {F.}~\bibnamefont {Krausz}},\ }\bibfield  {title} {\bibinfo
  {title} {{Attosecond Control and Measurement: Lightwave Electronics}},\
  }\href {https://doi.org/10.1126/science.1142855} {\bibfield  {journal}
  {\bibinfo  {journal} {Science}\ }\textbf {\bibinfo {volume} {317}},\ \bibinfo
  {pages} {769} (\bibinfo {year} {2007})}\BibitemShut {NoStop}%
\bibitem [{\citenamefont {Krausz}\ and\ \citenamefont
  {Stockman}(2014)}]{Krausz2014}%
  \BibitemOpen
  \bibfield  {author} {\bibinfo {author} {\bibfnamefont {F.}~\bibnamefont
  {Krausz}}\ and\ \bibinfo {author} {\bibfnamefont {M.~I.}\ \bibnamefont
  {Stockman}},\ }\bibfield  {title} {\bibinfo {title} {{Attosecond metrology:
  from electron capture to future signal processing}},\ }\href
  {https://doi.org/10.1038/nphoton.2014.28} {\bibfield  {journal} {\bibinfo
  {journal} {Nat. Photonics}\ }\textbf {\bibinfo {volume} {8}},\ \bibinfo
  {pages} {205} (\bibinfo {year} {2014})}\BibitemShut {NoStop}%
\bibitem [{\citenamefont {Luu}\ \emph {et~al.}(2015)\citenamefont {Luu},
  \citenamefont {Garg}, \citenamefont {Kruchinin}, \citenamefont {Moulet},
  \citenamefont {Hassan},\ and\ \citenamefont {Goulielmakis}}]{Luu2015a}%
  \BibitemOpen
  \bibfield  {author} {\bibinfo {author} {\bibfnamefont {T.~T.}\ \bibnamefont
  {Luu}}, \bibinfo {author} {\bibfnamefont {M.}~\bibnamefont {Garg}}, \bibinfo
  {author} {\bibfnamefont {S.~Y.}\ \bibnamefont {Kruchinin}}, \bibinfo {author}
  {\bibfnamefont {A.}~\bibnamefont {Moulet}}, \bibinfo {author} {\bibfnamefont
  {M.~T.}\ \bibnamefont {Hassan}},\ and\ \bibinfo {author} {\bibfnamefont
  {E.}~\bibnamefont {Goulielmakis}},\ }\bibfield  {title} {\bibinfo {title}
  {{Extreme ultraviolet high-harmonic spectroscopy of solids}},\ }\href
  {https://doi.org/10.1038/nature14456} {\bibfield  {journal} {\bibinfo
  {journal} {Nature}\ }\textbf {\bibinfo {volume} {521}},\ \bibinfo {pages}
  {498} (\bibinfo {year} {2015})}\BibitemShut {NoStop}%
\bibitem [{\citenamefont {Reimann}\ \emph {et~al.}(2018)\citenamefont
  {Reimann}, \citenamefont {Schlauderer}, \citenamefont {Schmid}, \citenamefont
  {Langer}, \citenamefont {Baierl}, \citenamefont {Kokh}, \citenamefont
  {Tereshchenko}, \citenamefont {Kimura}, \citenamefont {Lange}, \citenamefont
  {G{\"{u}}dde}, \citenamefont {H{\"{o}}fer},\ and\ \citenamefont
  {Huber}}]{Reimann2018}%
  \BibitemOpen
  \bibfield  {author} {\bibinfo {author} {\bibfnamefont {J.}~\bibnamefont
  {Reimann}}, \bibinfo {author} {\bibfnamefont {S.}~\bibnamefont
  {Schlauderer}}, \bibinfo {author} {\bibfnamefont {C.~P.}\ \bibnamefont
  {Schmid}}, \bibinfo {author} {\bibfnamefont {F.}~\bibnamefont {Langer}},
  \bibinfo {author} {\bibfnamefont {S.}~\bibnamefont {Baierl}}, \bibinfo
  {author} {\bibfnamefont {K.~A.}\ \bibnamefont {Kokh}}, \bibinfo {author}
  {\bibfnamefont {O.~E.}\ \bibnamefont {Tereshchenko}}, \bibinfo {author}
  {\bibfnamefont {A.}~\bibnamefont {Kimura}}, \bibinfo {author} {\bibfnamefont
  {C.}~\bibnamefont {Lange}}, \bibinfo {author} {\bibfnamefont
  {J.}~\bibnamefont {G{\"{u}}dde}}, \bibinfo {author} {\bibfnamefont
  {U.}~\bibnamefont {H{\"{o}}fer}},\ and\ \bibinfo {author} {\bibfnamefont
  {R.}~\bibnamefont {Huber}},\ }\bibfield  {title} {\bibinfo {title} {{Subcycle
  observation of lightwave-driven Dirac currents in a topological surface
  band}},\ }\href {https://doi.org/10.1038/s41586-018-0544-x} {\bibfield
  {journal} {\bibinfo  {journal} {Nature}\ }\textbf {\bibinfo {volume} {562}},\
  \bibinfo {pages} {396} (\bibinfo {year} {2018})}\BibitemShut {NoStop}%
\bibitem [{\citenamefont {Kawakami}\ \emph
  {et~al.}(2018{\natexlab{a}})\citenamefont {Kawakami}, \citenamefont {Itoh},
  \citenamefont {Yonemitsu},\ and\ \citenamefont {Iwai}}]{Kawakami2018}%
  \BibitemOpen
  \bibfield  {author} {\bibinfo {author} {\bibfnamefont {Y.}~\bibnamefont
  {Kawakami}}, \bibinfo {author} {\bibfnamefont {H.}~\bibnamefont {Itoh}},
  \bibinfo {author} {\bibfnamefont {K.}~\bibnamefont {Yonemitsu}},\ and\
  \bibinfo {author} {\bibfnamefont {S.}~\bibnamefont {Iwai}},\ }\bibfield
  {title} {\bibinfo {title} {{Strong light-field effects driven by nearly
  single-cycle 7 fs light-field in correlated organic conductors}},\ }\href
  {https://doi.org/10.1088/1361-6455/aad40a} {\bibfield  {journal} {\bibinfo
  {journal} {J. Phys. B}\ }\textbf {\bibinfo {volume} {51}},\ \bibinfo {pages}
  {174005} (\bibinfo {year} {2018}{\natexlab{a}})}\BibitemShut {NoStop}%
\bibitem [{\citenamefont {Kawakami}\ \emph
  {et~al.}(2018{\natexlab{b}})\citenamefont {Kawakami}, \citenamefont {Amano},
  \citenamefont {Yoneyama}, \citenamefont {Akamine}, \citenamefont {Itoh},
  \citenamefont {Kawaguchi}, \citenamefont {Yamamoto}, \citenamefont {Kishida},
  \citenamefont {Itoh}, \citenamefont {Sasaki}, \citenamefont {Ishihara},
  \citenamefont {Tanaka}, \citenamefont {Yonemitsu},\ and\ \citenamefont
  {Iwai}}]{Kawakami2018a}%
  \BibitemOpen
  \bibfield  {author} {\bibinfo {author} {\bibfnamefont {Y.}~\bibnamefont
  {Kawakami}}, \bibinfo {author} {\bibfnamefont {T.}~\bibnamefont {Amano}},
  \bibinfo {author} {\bibfnamefont {Y.}~\bibnamefont {Yoneyama}}, \bibinfo
  {author} {\bibfnamefont {Y.}~\bibnamefont {Akamine}}, \bibinfo {author}
  {\bibfnamefont {H.}~\bibnamefont {Itoh}}, \bibinfo {author} {\bibfnamefont
  {G.}~\bibnamefont {Kawaguchi}}, \bibinfo {author} {\bibfnamefont {H.~M.}\
  \bibnamefont {Yamamoto}}, \bibinfo {author} {\bibfnamefont {H.}~\bibnamefont
  {Kishida}}, \bibinfo {author} {\bibfnamefont {K.}~\bibnamefont {Itoh}},
  \bibinfo {author} {\bibfnamefont {T.}~\bibnamefont {Sasaki}}, \bibinfo
  {author} {\bibfnamefont {S.}~\bibnamefont {Ishihara}}, \bibinfo {author}
  {\bibfnamefont {Y.}~\bibnamefont {Tanaka}}, \bibinfo {author} {\bibfnamefont
  {K.}~\bibnamefont {Yonemitsu}},\ and\ \bibinfo {author} {\bibfnamefont
  {S.}~\bibnamefont {Iwai}},\ }\bibfield  {title} {\bibinfo {title} {{Nonlinear
  charge oscillation driven by a single-cycle light field in an organic
  superconductor}},\ }\href {https://doi.org/10.1038/s41566-018-0194-4}
  {\bibfield  {journal} {\bibinfo  {journal} {Nat. Photonics}\ }\textbf
  {\bibinfo {volume} {12}},\ \bibinfo {pages} {474} (\bibinfo {year}
  {2018}{\natexlab{b}})}\BibitemShut {NoStop}%
\bibitem [{\citenamefont {Schoetz}\ \emph {et~al.}(2019)\citenamefont
  {Schoetz}, \citenamefont {Wang}, \citenamefont {Pisanty}, \citenamefont
  {Lewenstein}, \citenamefont {Kling},\ and\ \citenamefont
  {Ciappina}}]{Schoetz2019a}%
  \BibitemOpen
  \bibfield  {author} {\bibinfo {author} {\bibfnamefont {J.}~\bibnamefont
  {Schoetz}}, \bibinfo {author} {\bibfnamefont {Z.}~\bibnamefont {Wang}},
  \bibinfo {author} {\bibfnamefont {E.}~\bibnamefont {Pisanty}}, \bibinfo
  {author} {\bibfnamefont {M.}~\bibnamefont {Lewenstein}}, \bibinfo {author}
  {\bibfnamefont {M.~F.}\ \bibnamefont {Kling}},\ and\ \bibinfo {author}
  {\bibfnamefont {M.~F.}\ \bibnamefont {Ciappina}},\ }\bibfield  {title}
  {\bibinfo {title} {{Perspective on Petahertz Electronics and Attosecond
  Nanoscopy}},\ }\href {https://doi.org/10.1021/acsphotonics.9b01188}
  {\bibfield  {journal} {\bibinfo  {journal} {ACS Photonics}\ }\textbf
  {\bibinfo {volume} {6}},\ \bibinfo {pages} {3057} (\bibinfo {year}
  {2019})}\BibitemShut {NoStop}%
\bibitem [{\citenamefont {Kawakami}\ \emph {et~al.}(2020)\citenamefont
  {Kawakami}, \citenamefont {Amano}, \citenamefont {Ohashi}, \citenamefont
  {Itoh}, \citenamefont {Nakamura}, \citenamefont {Kishida}, \citenamefont
  {Sasaki}, \citenamefont {Kawaguchi}, \citenamefont {Yamamoto}, \citenamefont
  {Yamamoto}, \citenamefont {Ishihara}, \citenamefont {Yonemitsu},\ and\
  \citenamefont {Iwai}}]{Kawakami2020}%
  \BibitemOpen
  \bibfield  {author} {\bibinfo {author} {\bibfnamefont {Y.}~\bibnamefont
  {Kawakami}}, \bibinfo {author} {\bibfnamefont {T.}~\bibnamefont {Amano}},
  \bibinfo {author} {\bibfnamefont {H.}~\bibnamefont {Ohashi}}, \bibinfo
  {author} {\bibfnamefont {H.}~\bibnamefont {Itoh}}, \bibinfo {author}
  {\bibfnamefont {Y.}~\bibnamefont {Nakamura}}, \bibinfo {author}
  {\bibfnamefont {H.}~\bibnamefont {Kishida}}, \bibinfo {author} {\bibfnamefont
  {T.}~\bibnamefont {Sasaki}}, \bibinfo {author} {\bibfnamefont
  {G.}~\bibnamefont {Kawaguchi}}, \bibinfo {author} {\bibfnamefont {H.~M.}\
  \bibnamefont {Yamamoto}}, \bibinfo {author} {\bibfnamefont {K.}~\bibnamefont
  {Yamamoto}}, \bibinfo {author} {\bibfnamefont {S.}~\bibnamefont {Ishihara}},
  \bibinfo {author} {\bibfnamefont {K.}~\bibnamefont {Yonemitsu}},\ and\
  \bibinfo {author} {\bibfnamefont {S.}~\bibnamefont {Iwai}},\ }\bibfield
  {title} {\bibinfo {title} {{Petahertz non-linear current in a centrosymmetric
  organic superconductor}},\ }\href
  {https://doi.org/10.1038/s41467-020-17776-3} {\bibfield  {journal} {\bibinfo
  {journal} {Nat. Commun.}\ }\textbf {\bibinfo {volume} {11}},\ \bibinfo
  {pages} {4138} (\bibinfo {year} {2020})}\BibitemShut {NoStop}%
\bibitem [{\citenamefont {Jim{\'{e}}nez-Gal{\'{a}}n}\ \emph
  {et~al.}(2021)\citenamefont {Jim{\'{e}}nez-Gal{\'{a}}n}, \citenamefont
  {Silva}, \citenamefont {Smirnova},\ and\ \citenamefont
  {Ivanov}}]{Jimenez-Galan2021}%
  \BibitemOpen
  \bibfield  {author} {\bibinfo {author} {\bibfnamefont {{\'{A}}.}~\bibnamefont
  {Jim{\'{e}}nez-Gal{\'{a}}n}}, \bibinfo {author} {\bibfnamefont {R.~E.~F.}\
  \bibnamefont {Silva}}, \bibinfo {author} {\bibfnamefont {O.}~\bibnamefont
  {Smirnova}},\ and\ \bibinfo {author} {\bibfnamefont {M.}~\bibnamefont
  {Ivanov}},\ }\bibfield  {title} {\bibinfo {title} {{Sub-cycle valleytronics:
  control of valley polarization using few-cycle linearly polarized pulses}},\
  }\href {https://doi.org/10.1364/OPTICA.404257} {\bibfield  {journal}
  {\bibinfo  {journal} {Optica}\ }\textbf {\bibinfo {volume} {8}},\ \bibinfo
  {pages} {277} (\bibinfo {year} {2021})}\BibitemShut {NoStop}%
\bibitem [{\citenamefont {Boolakee}\ \emph {et~al.}(2022)\citenamefont
  {Boolakee}, \citenamefont {Heide}, \citenamefont
  {Garz{\'{o}}n-Ram{\'{i}}rez}, \citenamefont {Weber}, \citenamefont {Franco},\
  and\ \citenamefont {Hommelhoff}}]{Boolakee2022a}%
  \BibitemOpen
  \bibfield  {author} {\bibinfo {author} {\bibfnamefont {T.}~\bibnamefont
  {Boolakee}}, \bibinfo {author} {\bibfnamefont {C.}~\bibnamefont {Heide}},
  \bibinfo {author} {\bibfnamefont {A.}~\bibnamefont
  {Garz{\'{o}}n-Ram{\'{i}}rez}}, \bibinfo {author} {\bibfnamefont {H.~B.}\
  \bibnamefont {Weber}}, \bibinfo {author} {\bibfnamefont {I.}~\bibnamefont
  {Franco}},\ and\ \bibinfo {author} {\bibfnamefont {P.}~\bibnamefont
  {Hommelhoff}},\ }\bibfield  {title} {\bibinfo {title} {{Light-field control
  of real and virtual charge carriers}},\ }\href
  {https://doi.org/10.1038/s41586-022-04565-9} {\bibfield  {journal} {\bibinfo
  {journal} {Nature}\ }\textbf {\bibinfo {volume} {605}},\ \bibinfo {pages}
  {251} (\bibinfo {year} {2022})}\BibitemShut {NoStop}%
\bibitem [{\citenamefont {McPherson}\ \emph {et~al.}(1987)\citenamefont
  {McPherson}, \citenamefont {Gibson}, \citenamefont {Jara}, \citenamefont
  {Johann}, \citenamefont {Luk}, \citenamefont {McIntyre}, \citenamefont
  {Boyer},\ and\ \citenamefont {Rhodes}}]{McPherson1987}%
  \BibitemOpen
  \bibfield  {author} {\bibinfo {author} {\bibfnamefont {A.}~\bibnamefont
  {McPherson}}, \bibinfo {author} {\bibfnamefont {G.}~\bibnamefont {Gibson}},
  \bibinfo {author} {\bibfnamefont {H.}~\bibnamefont {Jara}}, \bibinfo {author}
  {\bibfnamefont {U.}~\bibnamefont {Johann}}, \bibinfo {author} {\bibfnamefont
  {T.~S.}\ \bibnamefont {Luk}}, \bibinfo {author} {\bibfnamefont {I.~A.}\
  \bibnamefont {McIntyre}}, \bibinfo {author} {\bibfnamefont {K.}~\bibnamefont
  {Boyer}},\ and\ \bibinfo {author} {\bibfnamefont {C.~K.}\ \bibnamefont
  {Rhodes}},\ }\bibfield  {title} {\bibinfo {title} {{Studies of multiphoton
  production of vacuum-ultraviolet radiation in the rare gases}},\ }\href
  {https://doi.org/10.1364/JOSAB.4.000595} {\bibfield  {journal} {\bibinfo
  {journal} {J. Opt. Soc. Am. B}\ }\textbf {\bibinfo {volume} {4}},\ \bibinfo
  {pages} {595} (\bibinfo {year} {1987})}\BibitemShut {NoStop}%
\bibitem [{\citenamefont {Ferray}\ \emph {et~al.}(1988)\citenamefont {Ferray},
  \citenamefont {L'Huillier}, \citenamefont {Li}, \citenamefont {Lompre},
  \citenamefont {Mainfray},\ and\ \citenamefont {Manus}}]{Ferray1988}%
  \BibitemOpen
  \bibfield  {author} {\bibinfo {author} {\bibfnamefont {M.}~\bibnamefont
  {Ferray}}, \bibinfo {author} {\bibfnamefont {A.}~\bibnamefont {L'Huillier}},
  \bibinfo {author} {\bibfnamefont {X.~F.}\ \bibnamefont {Li}}, \bibinfo
  {author} {\bibfnamefont {L.~A.}\ \bibnamefont {Lompre}}, \bibinfo {author}
  {\bibfnamefont {G.}~\bibnamefont {Mainfray}},\ and\ \bibinfo {author}
  {\bibfnamefont {C.}~\bibnamefont {Manus}},\ }\bibfield  {title} {\bibinfo
  {title} {{Multiple-harmonic conversion of 1064 nm radiation in rare gases}},\
  }\href {https://doi.org/10.1088/0953-4075/21/3/001} {\bibfield  {journal}
  {\bibinfo  {journal} {J. Phys. B}\ }\textbf {\bibinfo {volume} {21}},\
  \bibinfo {pages} {L31} (\bibinfo {year} {1988})}\BibitemShut {NoStop}%
\bibitem [{\citenamefont {Corkum}(1993)}]{Corkum1993}%
  \BibitemOpen
  \bibfield  {author} {\bibinfo {author} {\bibfnamefont {P.~B.}\ \bibnamefont
  {Corkum}},\ }\bibfield  {title} {\bibinfo {title} {{Plasma perspective on
  strong field multiphoton ionization}},\ }\href
  {https://doi.org/10.1103/PhysRevLett.71.1994} {\bibfield  {journal} {\bibinfo
   {journal} {Phys. Rev. Lett.}\ }\textbf {\bibinfo {volume} {71}},\ \bibinfo
  {pages} {1994} (\bibinfo {year} {1993})}\BibitemShut {NoStop}%
\bibitem [{\citenamefont {Lewenstein}\ \emph {et~al.}(1994)\citenamefont
  {Lewenstein}, \citenamefont {Balcou}, \citenamefont {Ivanov}, \citenamefont
  {L'Huillier},\ and\ \citenamefont {Corkum}}]{Lewenstein1994}%
  \BibitemOpen
  \bibfield  {author} {\bibinfo {author} {\bibfnamefont {M.}~\bibnamefont
  {Lewenstein}}, \bibinfo {author} {\bibfnamefont {P.}~\bibnamefont {Balcou}},
  \bibinfo {author} {\bibfnamefont {M.~Y.}\ \bibnamefont {Ivanov}}, \bibinfo
  {author} {\bibfnamefont {A.}~\bibnamefont {L'Huillier}},\ and\ \bibinfo
  {author} {\bibfnamefont {P.~B.}\ \bibnamefont {Corkum}},\ }\bibfield  {title}
  {\bibinfo {title} {{Theory of high-harmonic generation by low-frequency laser
  fields}},\ }\href {https://doi.org/10.1103/PhysRevA.49.2117} {\bibfield
  {journal} {\bibinfo  {journal} {Phys. Rev. A}\ }\textbf {\bibinfo {volume}
  {49}},\ \bibinfo {pages} {2117} (\bibinfo {year} {1994})}\BibitemShut
  {NoStop}%
\bibitem [{\citenamefont {Hentschel}\ \emph {et~al.}(2001)\citenamefont
  {Hentschel}, \citenamefont {Kienberger}, \citenamefont {Spielmann},
  \citenamefont {Reider}, \citenamefont {Milosevic}, \citenamefont {Brabec},
  \citenamefont {Corkum}, \citenamefont {Heinzmann}, \citenamefont {Drescher},\
  and\ \citenamefont {Krausz}}]{Hentschel2001}%
  \BibitemOpen
  \bibfield  {author} {\bibinfo {author} {\bibfnamefont {M.}~\bibnamefont
  {Hentschel}}, \bibinfo {author} {\bibfnamefont {R.}~\bibnamefont
  {Kienberger}}, \bibinfo {author} {\bibfnamefont {C.}~\bibnamefont
  {Spielmann}}, \bibinfo {author} {\bibfnamefont {G.~A.}\ \bibnamefont
  {Reider}}, \bibinfo {author} {\bibfnamefont {N.}~\bibnamefont {Milosevic}},
  \bibinfo {author} {\bibfnamefont {T.}~\bibnamefont {Brabec}}, \bibinfo
  {author} {\bibfnamefont {P.}~\bibnamefont {Corkum}}, \bibinfo {author}
  {\bibfnamefont {U.}~\bibnamefont {Heinzmann}}, \bibinfo {author}
  {\bibfnamefont {M.}~\bibnamefont {Drescher}},\ and\ \bibinfo {author}
  {\bibfnamefont {F.}~\bibnamefont {Krausz}},\ }\bibfield  {title} {\bibinfo
  {title} {{Attosecond metrology}},\ }\href {https://doi.org/10.1038/35107000}
  {\bibfield  {journal} {\bibinfo  {journal} {Nature}\ }\textbf {\bibinfo
  {volume} {414}},\ \bibinfo {pages} {509} (\bibinfo {year}
  {2001})}\BibitemShut {NoStop}%
\bibitem [{\citenamefont {Ghimire}\ \emph {et~al.}(2011)\citenamefont
  {Ghimire}, \citenamefont {DiChiara}, \citenamefont {Sistrunk}, \citenamefont
  {Agostini}, \citenamefont {DiMauro},\ and\ \citenamefont
  {Reis}}]{Ghimire2011}%
  \BibitemOpen
  \bibfield  {author} {\bibinfo {author} {\bibfnamefont {S.}~\bibnamefont
  {Ghimire}}, \bibinfo {author} {\bibfnamefont {A.~D.}\ \bibnamefont
  {DiChiara}}, \bibinfo {author} {\bibfnamefont {E.}~\bibnamefont {Sistrunk}},
  \bibinfo {author} {\bibfnamefont {P.}~\bibnamefont {Agostini}}, \bibinfo
  {author} {\bibfnamefont {L.~F.}\ \bibnamefont {DiMauro}},\ and\ \bibinfo
  {author} {\bibfnamefont {D.~A.}\ \bibnamefont {Reis}},\ }\bibfield  {title}
  {\bibinfo {title} {{Observation of high-order harmonic generation in a bulk
  crystal}},\ }\href {https://doi.org/10.1038/nphys1847} {\bibfield  {journal}
  {\bibinfo  {journal} {Nat. Phys.}\ }\textbf {\bibinfo {volume} {7}},\
  \bibinfo {pages} {138} (\bibinfo {year} {2011})}\BibitemShut {NoStop}%
\bibitem [{\citenamefont {Schubert}\ \emph {et~al.}(2014)\citenamefont
  {Schubert}, \citenamefont {Hohenleutner}, \citenamefont {Langer},
  \citenamefont {Urbanek}, \citenamefont {Lange}, \citenamefont {Huttner},
  \citenamefont {Golde}, \citenamefont {Meier}, \citenamefont {Kira},
  \citenamefont {Koch},\ and\ \citenamefont {Huber}}]{Schubert2014a}%
  \BibitemOpen
  \bibfield  {author} {\bibinfo {author} {\bibfnamefont {O.}~\bibnamefont
  {Schubert}}, \bibinfo {author} {\bibfnamefont {M.}~\bibnamefont
  {Hohenleutner}}, \bibinfo {author} {\bibfnamefont {F.}~\bibnamefont
  {Langer}}, \bibinfo {author} {\bibfnamefont {B.}~\bibnamefont {Urbanek}},
  \bibinfo {author} {\bibfnamefont {C.}~\bibnamefont {Lange}}, \bibinfo
  {author} {\bibfnamefont {U.}~\bibnamefont {Huttner}}, \bibinfo {author}
  {\bibfnamefont {D.}~\bibnamefont {Golde}}, \bibinfo {author} {\bibfnamefont
  {T.}~\bibnamefont {Meier}}, \bibinfo {author} {\bibfnamefont
  {M.}~\bibnamefont {Kira}}, \bibinfo {author} {\bibfnamefont {S.~W.}\
  \bibnamefont {Koch}},\ and\ \bibinfo {author} {\bibfnamefont
  {R.}~\bibnamefont {Huber}},\ }\bibfield  {title} {\bibinfo {title}
  {{Sub-cycle control of terahertz high-harmonic generation by dynamical Bloch
  oscillations}},\ }\href {https://doi.org/10.1038/nphoton.2013.349} {\bibfield
   {journal} {\bibinfo  {journal} {Nat. Photonics}\ }\textbf {\bibinfo {volume}
  {8}},\ \bibinfo {pages} {119} (\bibinfo {year} {2014})}\BibitemShut {NoStop}%
\bibitem [{\citenamefont {Hohenleutner}\ \emph {et~al.}(2015)\citenamefont
  {Hohenleutner}, \citenamefont {Langer}, \citenamefont {Schubert},
  \citenamefont {Knorr}, \citenamefont {Huttner}, \citenamefont {Koch},
  \citenamefont {Kira},\ and\ \citenamefont {Huber}}]{Hohenleutner2015}%
  \BibitemOpen
  \bibfield  {author} {\bibinfo {author} {\bibfnamefont {M.}~\bibnamefont
  {Hohenleutner}}, \bibinfo {author} {\bibfnamefont {F.}~\bibnamefont
  {Langer}}, \bibinfo {author} {\bibfnamefont {O.}~\bibnamefont {Schubert}},
  \bibinfo {author} {\bibfnamefont {M.}~\bibnamefont {Knorr}}, \bibinfo
  {author} {\bibfnamefont {U.}~\bibnamefont {Huttner}}, \bibinfo {author}
  {\bibfnamefont {S.~W.}\ \bibnamefont {Koch}}, \bibinfo {author}
  {\bibfnamefont {M.}~\bibnamefont {Kira}},\ and\ \bibinfo {author}
  {\bibfnamefont {R.}~\bibnamefont {Huber}},\ }\bibfield  {title} {\bibinfo
  {title} {{Real-time observation of interfering crystal electrons in
  high-harmonic generation}},\ }\href {https://doi.org/10.1038/nature14652}
  {\bibfield  {journal} {\bibinfo  {journal} {Nature}\ }\textbf {\bibinfo
  {volume} {523}},\ \bibinfo {pages} {572} (\bibinfo {year}
  {2015})}\BibitemShut {NoStop}%
\bibitem [{\citenamefont {Garg}\ \emph
  {et~al.}(2016{\natexlab{a}})\citenamefont {Garg}, \citenamefont {Zhan},
  \citenamefont {Luu}, \citenamefont {Lakhotia}, \citenamefont {Klostermann},
  \citenamefont {Guggenmos},\ and\ \citenamefont {Goulielmakis}}]{Garg2016a}%
  \BibitemOpen
  \bibfield  {author} {\bibinfo {author} {\bibfnamefont {M.}~\bibnamefont
  {Garg}}, \bibinfo {author} {\bibfnamefont {M.}~\bibnamefont {Zhan}}, \bibinfo
  {author} {\bibfnamefont {T.~T.}\ \bibnamefont {Luu}}, \bibinfo {author}
  {\bibfnamefont {H.}~\bibnamefont {Lakhotia}}, \bibinfo {author}
  {\bibfnamefont {T.}~\bibnamefont {Klostermann}}, \bibinfo {author}
  {\bibfnamefont {A.}~\bibnamefont {Guggenmos}},\ and\ \bibinfo {author}
  {\bibfnamefont {E.}~\bibnamefont {Goulielmakis}},\ }\bibfield  {title}
  {\bibinfo {title} {{Multi-petahertz electronic metrology}},\ }\href
  {https://doi.org/10.1038/nature19821} {\bibfield  {journal} {\bibinfo
  {journal} {Nature}\ }\textbf {\bibinfo {volume} {538}},\ \bibinfo {pages}
  {359} (\bibinfo {year} {2016}{\natexlab{a}})}\BibitemShut {NoStop}%
\bibitem [{\citenamefont {Langer}\ \emph {et~al.}(2016)\citenamefont {Langer},
  \citenamefont {Hohenleutner}, \citenamefont {Schmid}, \citenamefont
  {Poellmann}, \citenamefont {Nagler}, \citenamefont {Korn}, \citenamefont
  {Sch{\"{u}}ller}, \citenamefont {Sherwin}, \citenamefont {Huttner},
  \citenamefont {Steiner}, \citenamefont {Koch}, \citenamefont {Kira},\ and\
  \citenamefont {Huber}}]{Langer2016}%
  \BibitemOpen
  \bibfield  {author} {\bibinfo {author} {\bibfnamefont {F.}~\bibnamefont
  {Langer}}, \bibinfo {author} {\bibfnamefont {M.}~\bibnamefont
  {Hohenleutner}}, \bibinfo {author} {\bibfnamefont {C.~P.}\ \bibnamefont
  {Schmid}}, \bibinfo {author} {\bibfnamefont {C.}~\bibnamefont {Poellmann}},
  \bibinfo {author} {\bibfnamefont {P.}~\bibnamefont {Nagler}}, \bibinfo
  {author} {\bibfnamefont {T.}~\bibnamefont {Korn}}, \bibinfo {author}
  {\bibfnamefont {C.}~\bibnamefont {Sch{\"{u}}ller}}, \bibinfo {author}
  {\bibfnamefont {M.~S.}\ \bibnamefont {Sherwin}}, \bibinfo {author}
  {\bibfnamefont {U.}~\bibnamefont {Huttner}}, \bibinfo {author} {\bibfnamefont
  {J.~T.}\ \bibnamefont {Steiner}}, \bibinfo {author} {\bibfnamefont {S.~W.}\
  \bibnamefont {Koch}}, \bibinfo {author} {\bibfnamefont {M.}~\bibnamefont
  {Kira}},\ and\ \bibinfo {author} {\bibfnamefont {R.}~\bibnamefont {Huber}},\
  }\bibfield  {title} {\bibinfo {title} {{Lightwave-driven quasiparticle
  collisions on a subcycle timescale}},\ }\href
  {https://doi.org/10.1038/nature17958} {\bibfield  {journal} {\bibinfo
  {journal} {Nature}\ }\textbf {\bibinfo {volume} {533}},\ \bibinfo {pages}
  {225} (\bibinfo {year} {2016})}\BibitemShut {NoStop}%
\bibitem [{\citenamefont {Langer}\ \emph {et~al.}(2017)\citenamefont {Langer},
  \citenamefont {Hohenleutner}, \citenamefont {Huttner}, \citenamefont {Koch},
  \citenamefont {Kira},\ and\ \citenamefont {Huber}}]{Langer2017}%
  \BibitemOpen
  \bibfield  {author} {\bibinfo {author} {\bibfnamefont {F.}~\bibnamefont
  {Langer}}, \bibinfo {author} {\bibfnamefont {M.}~\bibnamefont
  {Hohenleutner}}, \bibinfo {author} {\bibfnamefont {U.}~\bibnamefont
  {Huttner}}, \bibinfo {author} {\bibfnamefont {S.~W.}\ \bibnamefont {Koch}},
  \bibinfo {author} {\bibfnamefont {M.}~\bibnamefont {Kira}},\ and\ \bibinfo
  {author} {\bibfnamefont {R.}~\bibnamefont {Huber}},\ }\bibfield  {title}
  {\bibinfo {title} {{Symmetry-controlled temporal structure of high-harmonic
  carrier fields from a bulk crystal}},\ }\href
  {https://doi.org/10.1038/nphoton.2017.29} {\bibfield  {journal} {\bibinfo
  {journal} {Nat. Photonics}\ }\textbf {\bibinfo {volume} {11}},\ \bibinfo
  {pages} {227} (\bibinfo {year} {2017})}\BibitemShut {NoStop}%
\bibitem [{\citenamefont {Huttner}\ \emph {et~al.}(2017)\citenamefont
  {Huttner}, \citenamefont {Kira},\ and\ \citenamefont {Koch}}]{Huttner2017}%
  \BibitemOpen
  \bibfield  {author} {\bibinfo {author} {\bibfnamefont {U.}~\bibnamefont
  {Huttner}}, \bibinfo {author} {\bibfnamefont {M.}~\bibnamefont {Kira}},\ and\
  \bibinfo {author} {\bibfnamefont {S.~W.}\ \bibnamefont {Koch}},\ }\bibfield
  {title} {\bibinfo {title} {{Ultrahigh Off-Resonant Field Effects in
  Semiconductors}},\ }\href {https://doi.org/10.1002/lpor.201700049} {\bibfield
   {journal} {\bibinfo  {journal} {Laser Photon. Rev.}\ }\textbf {\bibinfo
  {volume} {11}},\ \bibinfo {pages} {1700049} (\bibinfo {year}
  {2017})}\BibitemShut {NoStop}%
\bibitem [{\citenamefont {Kruchinin}\ \emph {et~al.}(2018)\citenamefont
  {Kruchinin}, \citenamefont {Krausz},\ and\ \citenamefont
  {Yakovlev}}]{Kruchinin2018}%
  \BibitemOpen
  \bibfield  {author} {\bibinfo {author} {\bibfnamefont {S.~Y.}\ \bibnamefont
  {Kruchinin}}, \bibinfo {author} {\bibfnamefont {F.}~\bibnamefont {Krausz}},\
  and\ \bibinfo {author} {\bibfnamefont {V.~S.}\ \bibnamefont {Yakovlev}},\
  }\bibfield  {title} {\bibinfo {title} {{Colloquium: Strong-field phenomena in
  periodic systems}},\ }\href {https://doi.org/10.1103/RevModPhys.90.021002}
  {\bibfield  {journal} {\bibinfo  {journal} {Rev. Mod. Phys.}\ }\textbf
  {\bibinfo {volume} {90}},\ \bibinfo {pages} {021002} (\bibinfo {year}
  {2018})}\BibitemShut {NoStop}%
\bibitem [{\citenamefont {Ortmann}\ and\ \citenamefont
  {Landsman}(2021)}]{Ortmann2021}%
  \BibitemOpen
  \bibfield  {author} {\bibinfo {author} {\bibfnamefont {L.}~\bibnamefont
  {Ortmann}}\ and\ \bibinfo {author} {\bibfnamefont {A.~S.}\ \bibnamefont
  {Landsman}},\ }\bibfield  {title} {\bibinfo {title} {{High-harmonic
  generation in solids}},\ }in\ \href
  {https://doi.org/10.1016/bs.aamop.2021.04.002} {\emph {\bibinfo {booktitle}
  {Adv. At. Mol. Opt. Phys.}}},\ Vol.~\bibinfo {volume} {70}\ (\bibinfo
  {publisher} {Elsevier},\ \bibinfo {year} {2021})\ \bibinfo {edition} {1st}\
  ed.,\ pp.\ \bibinfo {pages} {103--156}\BibitemShut {NoStop}%
\bibitem [{\citenamefont {Xia}\ \emph {et~al.}(2021)\citenamefont {Xia},
  \citenamefont {Tamaya}, \citenamefont {Kim}, \citenamefont {Lu},
  \citenamefont {Kanai}, \citenamefont {Ishii}, \citenamefont {Itatani},
  \citenamefont {Akiyama},\ and\ \citenamefont {Kato}}]{Xia2021}%
  \BibitemOpen
  \bibfield  {author} {\bibinfo {author} {\bibfnamefont {P.}~\bibnamefont
  {Xia}}, \bibinfo {author} {\bibfnamefont {T.}~\bibnamefont {Tamaya}},
  \bibinfo {author} {\bibfnamefont {C.}~\bibnamefont {Kim}}, \bibinfo {author}
  {\bibfnamefont {F.}~\bibnamefont {Lu}}, \bibinfo {author} {\bibfnamefont
  {T.}~\bibnamefont {Kanai}}, \bibinfo {author} {\bibfnamefont
  {N.}~\bibnamefont {Ishii}}, \bibinfo {author} {\bibfnamefont
  {J.}~\bibnamefont {Itatani}}, \bibinfo {author} {\bibfnamefont
  {H.}~\bibnamefont {Akiyama}},\ and\ \bibinfo {author} {\bibfnamefont
  {T.}~\bibnamefont {Kato}},\ }\bibfield  {title} {\bibinfo {title}
  {{High-harmonic generation in GaAs beyond the perturbative regime}},\ }\href
  {https://doi.org/10.1103/PhysRevB.104.L121202} {\bibfield  {journal}
  {\bibinfo  {journal} {Phys. Rev. B}\ }\textbf {\bibinfo {volume} {104}},\
  \bibinfo {pages} {L121202} (\bibinfo {year} {2021})}\BibitemShut {NoStop}%
\bibitem [{\citenamefont {Tamaya}\ \emph {et~al.}(2016)\citenamefont {Tamaya},
  \citenamefont {Ishikawa}, \citenamefont {Ogawa},\ and\ \citenamefont
  {Tanaka}}]{Tamaya2016}%
  \BibitemOpen
  \bibfield  {author} {\bibinfo {author} {\bibfnamefont {T.}~\bibnamefont
  {Tamaya}}, \bibinfo {author} {\bibfnamefont {A.}~\bibnamefont {Ishikawa}},
  \bibinfo {author} {\bibfnamefont {T.}~\bibnamefont {Ogawa}},\ and\ \bibinfo
  {author} {\bibfnamefont {K.}~\bibnamefont {Tanaka}},\ }\bibfield  {title}
  {\bibinfo {title} {{Diabatic Mechanisms of Higher-Order Harmonic Generation
  in Solid-State Materials under High-Intensity Electric Fields}},\ }\href
  {https://doi.org/10.1103/PhysRevLett.116.016601} {\bibfield  {journal}
  {\bibinfo  {journal} {Phys. Rev. Lett.}\ }\textbf {\bibinfo {volume} {116}},\
  \bibinfo {pages} {016601} (\bibinfo {year} {2016})}\BibitemShut {NoStop}%
\bibitem [{\citenamefont {Yoshikawa}\ \emph {et~al.}(2017)\citenamefont
  {Yoshikawa}, \citenamefont {Tamaya},\ and\ \citenamefont
  {Tanaka}}]{Yoshikawa2017}%
  \BibitemOpen
  \bibfield  {author} {\bibinfo {author} {\bibfnamefont {N.}~\bibnamefont
  {Yoshikawa}}, \bibinfo {author} {\bibfnamefont {T.}~\bibnamefont {Tamaya}},\
  and\ \bibinfo {author} {\bibfnamefont {K.}~\bibnamefont {Tanaka}},\
  }\bibfield  {title} {\bibinfo {title} {{High-harmonic generation in graphene
  enhanced by elliptically polarized light excitation}},\ }\href
  {https://doi.org/10.1126/science.aam8861} {\bibfield  {journal} {\bibinfo
  {journal} {Science}\ }\textbf {\bibinfo {volume} {356}},\ \bibinfo {pages}
  {736} (\bibinfo {year} {2017})}\BibitemShut {NoStop}%
\bibitem [{\citenamefont {Bai}\ \emph {et~al.}(2021)\citenamefont {Bai},
  \citenamefont {Fei}, \citenamefont {Wang}, \citenamefont {Li}, \citenamefont
  {Li}, \citenamefont {Song}, \citenamefont {Li}, \citenamefont {Xu},\ and\
  \citenamefont {Liu}}]{Bai2020a}%
  \BibitemOpen
  \bibfield  {author} {\bibinfo {author} {\bibfnamefont {Y.}~\bibnamefont
  {Bai}}, \bibinfo {author} {\bibfnamefont {F.}~\bibnamefont {Fei}}, \bibinfo
  {author} {\bibfnamefont {S.}~\bibnamefont {Wang}}, \bibinfo {author}
  {\bibfnamefont {N.}~\bibnamefont {Li}}, \bibinfo {author} {\bibfnamefont
  {X.}~\bibnamefont {Li}}, \bibinfo {author} {\bibfnamefont {F.}~\bibnamefont
  {Song}}, \bibinfo {author} {\bibfnamefont {R.}~\bibnamefont {Li}}, \bibinfo
  {author} {\bibfnamefont {Z.}~\bibnamefont {Xu}},\ and\ \bibinfo {author}
  {\bibfnamefont {P.}~\bibnamefont {Liu}},\ }\bibfield  {title} {\bibinfo
  {title} {{High-harmonic generation from topological surface states}},\ }\href
  {https://doi.org/10.1038/s41567-020-01052-8} {\bibfield  {journal} {\bibinfo
  {journal} {Nat. Phys.}\ }\textbf {\bibinfo {volume} {17}},\ \bibinfo {pages}
  {311} (\bibinfo {year} {2021})}\BibitemShut {NoStop}%
\bibitem [{\citenamefont {Schmid}\ \emph {et~al.}(2021)\citenamefont {Schmid},
  \citenamefont {Weigl}, \citenamefont {Gr{\"{o}}ssing}, \citenamefont {Junk},
  \citenamefont {Gorini}, \citenamefont {Schlauderer}, \citenamefont {Ito},
  \citenamefont {Meierhofer}, \citenamefont {Hofmann}, \citenamefont
  {Afanasiev}, \citenamefont {Crewse}, \citenamefont {Kokh}, \citenamefont
  {Tereshchenko}, \citenamefont {G{\"{u}}dde}, \citenamefont {Evers},
  \citenamefont {Wilhelm}, \citenamefont {Richter}, \citenamefont
  {H{\"{o}}fer},\ and\ \citenamefont {Huber}}]{Schmid2021}%
  \BibitemOpen
  \bibfield  {author} {\bibinfo {author} {\bibfnamefont {C.~P.}\ \bibnamefont
  {Schmid}}, \bibinfo {author} {\bibfnamefont {L.}~\bibnamefont {Weigl}},
  \bibinfo {author} {\bibfnamefont {P.}~\bibnamefont {Gr{\"{o}}ssing}},
  \bibinfo {author} {\bibfnamefont {V.}~\bibnamefont {Junk}}, \bibinfo {author}
  {\bibfnamefont {C.}~\bibnamefont {Gorini}}, \bibinfo {author} {\bibfnamefont
  {S.}~\bibnamefont {Schlauderer}}, \bibinfo {author} {\bibfnamefont
  {S.}~\bibnamefont {Ito}}, \bibinfo {author} {\bibfnamefont {M.}~\bibnamefont
  {Meierhofer}}, \bibinfo {author} {\bibfnamefont {N.}~\bibnamefont {Hofmann}},
  \bibinfo {author} {\bibfnamefont {D.}~\bibnamefont {Afanasiev}}, \bibinfo
  {author} {\bibfnamefont {J.}~\bibnamefont {Crewse}}, \bibinfo {author}
  {\bibfnamefont {K.~A.}\ \bibnamefont {Kokh}}, \bibinfo {author}
  {\bibfnamefont {O.~E.}\ \bibnamefont {Tereshchenko}}, \bibinfo {author}
  {\bibfnamefont {J.}~\bibnamefont {G{\"{u}}dde}}, \bibinfo {author}
  {\bibfnamefont {F.}~\bibnamefont {Evers}}, \bibinfo {author} {\bibfnamefont
  {J.}~\bibnamefont {Wilhelm}}, \bibinfo {author} {\bibfnamefont
  {K.}~\bibnamefont {Richter}}, \bibinfo {author} {\bibfnamefont
  {U.}~\bibnamefont {H{\"{o}}fer}},\ and\ \bibinfo {author} {\bibfnamefont
  {R.}~\bibnamefont {Huber}},\ }\bibfield  {title} {\bibinfo {title} {{Tunable
  non-integer high-harmonic generation in a topological insulator}},\ }\href
  {https://doi.org/10.1038/s41586-021-03466-7} {\bibfield  {journal} {\bibinfo
  {journal} {Nature}\ }\textbf {\bibinfo {volume} {593}},\ \bibinfo {pages}
  {385} (\bibinfo {year} {2021})}\BibitemShut {NoStop}%
\bibitem [{\citenamefont {Lv}\ \emph {et~al.}(2021)\citenamefont {Lv},
  \citenamefont {Xu}, \citenamefont {Han}, \citenamefont {Zhang}, \citenamefont
  {Han}, \citenamefont {Zhou}, \citenamefont {Yao}, \citenamefont {Liu},
  \citenamefont {Lu}, \citenamefont {Weng}, \citenamefont {Xie}, \citenamefont
  {Chen}, \citenamefont {Hu}, \citenamefont {Chen},\ and\ \citenamefont
  {Zhu}}]{Lv2021}%
  \BibitemOpen
  \bibfield  {author} {\bibinfo {author} {\bibfnamefont {Y.-y.}\ \bibnamefont
  {Lv}}, \bibinfo {author} {\bibfnamefont {J.}~\bibnamefont {Xu}}, \bibinfo
  {author} {\bibfnamefont {S.}~\bibnamefont {Han}}, \bibinfo {author}
  {\bibfnamefont {C.}~\bibnamefont {Zhang}}, \bibinfo {author} {\bibfnamefont
  {Y.}~\bibnamefont {Han}}, \bibinfo {author} {\bibfnamefont {J.}~\bibnamefont
  {Zhou}}, \bibinfo {author} {\bibfnamefont {S.-h.}\ \bibnamefont {Yao}},
  \bibinfo {author} {\bibfnamefont {X.-p.}\ \bibnamefont {Liu}}, \bibinfo
  {author} {\bibfnamefont {M.-h.}\ \bibnamefont {Lu}}, \bibinfo {author}
  {\bibfnamefont {H.}~\bibnamefont {Weng}}, \bibinfo {author} {\bibfnamefont
  {Z.}~\bibnamefont {Xie}}, \bibinfo {author} {\bibfnamefont {Y.~B.}\
  \bibnamefont {Chen}}, \bibinfo {author} {\bibfnamefont {J.}~\bibnamefont
  {Hu}}, \bibinfo {author} {\bibfnamefont {Y.-f.}\ \bibnamefont {Chen}},\ and\
  \bibinfo {author} {\bibfnamefont {S.}~\bibnamefont {Zhu}},\ }\bibfield
  {title} {\bibinfo {title} {{High-harmonic generation in Weyl semimetal
  $\beta$-$\mathrm{WP_2}$ crystals}},\ }\href
  {https://doi.org/10.1038/s41467-021-26766-y} {\bibfield  {journal} {\bibinfo
  {journal} {Nat. Commun.}\ }\textbf {\bibinfo {volume} {12}},\ \bibinfo
  {pages} {6437} (\bibinfo {year} {2021})}\BibitemShut {NoStop}%
\bibitem [{\citenamefont {Silva}\ \emph {et~al.}(2018)\citenamefont {Silva},
  \citenamefont {Blinov}, \citenamefont {Rubtsov}, \citenamefont {Smirnova},\
  and\ \citenamefont {Ivanov}}]{Silva2018}%
  \BibitemOpen
  \bibfield  {author} {\bibinfo {author} {\bibfnamefont {R.~E.~F.}\
  \bibnamefont {Silva}}, \bibinfo {author} {\bibfnamefont {I.~V.}\ \bibnamefont
  {Blinov}}, \bibinfo {author} {\bibfnamefont {A.~N.}\ \bibnamefont {Rubtsov}},
  \bibinfo {author} {\bibfnamefont {O.}~\bibnamefont {Smirnova}},\ and\
  \bibinfo {author} {\bibfnamefont {M.}~\bibnamefont {Ivanov}},\ }\bibfield
  {title} {\bibinfo {title} {{High-harmonic spectroscopy of ultrafast many-body
  dynamics in strongly correlated systems}},\ }\href
  {https://doi.org/10.1038/s41566-018-0129-0} {\bibfield  {journal} {\bibinfo
  {journal} {Nat. Photonics}\ }\textbf {\bibinfo {volume} {12}},\ \bibinfo
  {pages} {266} (\bibinfo {year} {2018})}\BibitemShut {NoStop}%
\bibitem [{\citenamefont {Murakami}\ \emph {et~al.}(2018)\citenamefont
  {Murakami}, \citenamefont {Eckstein},\ and\ \citenamefont
  {Werner}}]{Murakami2018d}%
  \BibitemOpen
  \bibfield  {author} {\bibinfo {author} {\bibfnamefont {Y.}~\bibnamefont
  {Murakami}}, \bibinfo {author} {\bibfnamefont {M.}~\bibnamefont {Eckstein}},\
  and\ \bibinfo {author} {\bibfnamefont {P.}~\bibnamefont {Werner}},\
  }\bibfield  {title} {\bibinfo {title} {{High-Harmonic Generation in Mott
  Insulators}},\ }\href {https://doi.org/10.1103/PhysRevLett.121.057405}
  {\bibfield  {journal} {\bibinfo  {journal} {Phys. Rev. Lett.}\ }\textbf
  {\bibinfo {volume} {121}},\ \bibinfo {pages} {057405} (\bibinfo {year}
  {2018})}\BibitemShut {NoStop}%
\bibitem [{\citenamefont {Tancogne-Dejean}\ \emph {et~al.}(2018)\citenamefont
  {Tancogne-Dejean}, \citenamefont {Sentef},\ and\ \citenamefont
  {Rubio}}]{Tancogne-Dejean2018a}%
  \BibitemOpen
  \bibfield  {author} {\bibinfo {author} {\bibfnamefont {N.}~\bibnamefont
  {Tancogne-Dejean}}, \bibinfo {author} {\bibfnamefont {M.~A.}\ \bibnamefont
  {Sentef}},\ and\ \bibinfo {author} {\bibfnamefont {A.}~\bibnamefont
  {Rubio}},\ }\bibfield  {title} {\bibinfo {title} {{Ultrafast Modification of
  Hubbard $U$ in a Strongly Correlated Material: Ab initio High-Harmonic
  Generation in NiO}},\ }\href {https://doi.org/10.1103/PhysRevLett.121.097402}
  {\bibfield  {journal} {\bibinfo  {journal} {Phys. Rev. Lett.}\ }\textbf
  {\bibinfo {volume} {121}},\ \bibinfo {pages} {097402} (\bibinfo {year}
  {2018})}\BibitemShut {NoStop}%
\bibitem [{\citenamefont {Nag}\ \emph {et~al.}(2019)\citenamefont {Nag},
  \citenamefont {Slager}, \citenamefont {Higuchi},\ and\ \citenamefont
  {Oka}}]{Nag2019}%
  \BibitemOpen
  \bibfield  {author} {\bibinfo {author} {\bibfnamefont {T.}~\bibnamefont
  {Nag}}, \bibinfo {author} {\bibfnamefont {R.-J.}\ \bibnamefont {Slager}},
  \bibinfo {author} {\bibfnamefont {T.}~\bibnamefont {Higuchi}},\ and\ \bibinfo
  {author} {\bibfnamefont {T.}~\bibnamefont {Oka}},\ }\bibfield  {title}
  {\bibinfo {title} {{Dynamical synchronization transition in interacting
  electron systems}},\ }\href {https://doi.org/10.1103/PhysRevB.100.134301}
  {\bibfield  {journal} {\bibinfo  {journal} {Phys. Rev. B}\ }\textbf {\bibinfo
  {volume} {100}},\ \bibinfo {pages} {134301} (\bibinfo {year}
  {2019})}\BibitemShut {NoStop}%
\bibitem [{\citenamefont {Roy}\ \emph {et~al.}(2020)\citenamefont {Roy},
  \citenamefont {Bera},\ and\ \citenamefont {Saha}}]{Roy2020a}%
  \BibitemOpen
  \bibfield  {author} {\bibinfo {author} {\bibfnamefont {A.}~\bibnamefont
  {Roy}}, \bibinfo {author} {\bibfnamefont {S.}~\bibnamefont {Bera}},\ and\
  \bibinfo {author} {\bibfnamefont {K.}~\bibnamefont {Saha}},\ }\bibfield
  {title} {\bibinfo {title} {{Nonlinear dynamical response of interacting
  bosons to synthetic electric field}},\ }\href
  {https://doi.org/10.1103/PhysRevResearch.2.043133} {\bibfield  {journal}
  {\bibinfo  {journal} {Phys. Rev. Res.}\ }\textbf {\bibinfo {volume} {2}},\
  \bibinfo {pages} {043133} (\bibinfo {year} {2020})}\BibitemShut {NoStop}%
\bibitem [{\citenamefont {Murakami}\ \emph {et~al.}(2021)\citenamefont
  {Murakami}, \citenamefont {Takayoshi}, \citenamefont {Koga},\ and\
  \citenamefont {Werner}}]{Murakami2021}%
  \BibitemOpen
  \bibfield  {author} {\bibinfo {author} {\bibfnamefont {Y.}~\bibnamefont
  {Murakami}}, \bibinfo {author} {\bibfnamefont {S.}~\bibnamefont {Takayoshi}},
  \bibinfo {author} {\bibfnamefont {A.}~\bibnamefont {Koga}},\ and\ \bibinfo
  {author} {\bibfnamefont {P.}~\bibnamefont {Werner}},\ }\bibfield  {title}
  {\bibinfo {title} {{High-harmonic generation in one-dimensional Mott
  insulators}},\ }\href {https://doi.org/10.1103/PhysRevB.103.035110}
  {\bibfield  {journal} {\bibinfo  {journal} {Phys. Rev. B}\ }\textbf {\bibinfo
  {volume} {103}},\ \bibinfo {pages} {035110} (\bibinfo {year}
  {2021})}\BibitemShut {NoStop}%
\bibitem [{\citenamefont {Orthodoxou}\ \emph {et~al.}(2021)\citenamefont
  {Orthodoxou}, \citenamefont {Za{\"{i}}r},\ and\ \citenamefont
  {Booth}}]{Orthodoxou2021a}%
  \BibitemOpen
  \bibfield  {author} {\bibinfo {author} {\bibfnamefont {C.}~\bibnamefont
  {Orthodoxou}}, \bibinfo {author} {\bibfnamefont {A.}~\bibnamefont
  {Za{\"{i}}r}},\ and\ \bibinfo {author} {\bibfnamefont {G.~H.}\ \bibnamefont
  {Booth}},\ }\bibfield  {title} {\bibinfo {title} {{High harmonic generation
  in two-dimensional Mott insulators}},\ }\href
  {https://doi.org/10.1038/s41535-021-00377-8} {\bibfield  {journal} {\bibinfo
  {journal} {npj Quantum Mater.}\ }\textbf {\bibinfo {volume} {6}},\ \bibinfo
  {pages} {76} (\bibinfo {year} {2021})}\BibitemShut {NoStop}%
\bibitem [{\citenamefont {Udono}\ \emph {et~al.}(2022)\citenamefont {Udono},
  \citenamefont {Sugimoto}, \citenamefont {Kaneko},\ and\ \citenamefont
  {Ohta}}]{Udono2022a}%
  \BibitemOpen
  \bibfield  {author} {\bibinfo {author} {\bibfnamefont {M.}~\bibnamefont
  {Udono}}, \bibinfo {author} {\bibfnamefont {K.}~\bibnamefont {Sugimoto}},
  \bibinfo {author} {\bibfnamefont {T.}~\bibnamefont {Kaneko}},\ and\ \bibinfo
  {author} {\bibfnamefont {Y.}~\bibnamefont {Ohta}},\ }\bibfield  {title}
  {\bibinfo {title} {{Excitonic effects on high-harmonic generation in Mott
  insulators}},\ }\href {https://doi.org/10.1103/PhysRevB.105.L241108}
  {\bibfield  {journal} {\bibinfo  {journal} {Phys. Rev. B}\ }\textbf {\bibinfo
  {volume} {105}},\ \bibinfo {pages} {L241108} (\bibinfo {year}
  {2022})}\BibitemShut {NoStop}%
\bibitem [{\citenamefont {Shao}\ \emph {et~al.}(2022)\citenamefont {Shao},
  \citenamefont {Lu}, \citenamefont {Zhang}, \citenamefont {Yu}, \citenamefont
  {Tohyama},\ and\ \citenamefont {Lu}}]{Shao2022}%
  \BibitemOpen
  \bibfield  {author} {\bibinfo {author} {\bibfnamefont {C.}~\bibnamefont
  {Shao}}, \bibinfo {author} {\bibfnamefont {H.}~\bibnamefont {Lu}}, \bibinfo
  {author} {\bibfnamefont {X.}~\bibnamefont {Zhang}}, \bibinfo {author}
  {\bibfnamefont {C.}~\bibnamefont {Yu}}, \bibinfo {author} {\bibfnamefont
  {T.}~\bibnamefont {Tohyama}},\ and\ \bibinfo {author} {\bibfnamefont
  {R.}~\bibnamefont {Lu}},\ }\bibfield  {title} {\bibinfo {title}
  {{High-Harmonic Generation Approaching the Quantum Critical Point of Strongly
  Correlated Systems}},\ }\href
  {https://doi.org/10.1103/PhysRevLett.128.047401} {\bibfield  {journal}
  {\bibinfo  {journal} {Phys. Rev. Lett.}\ }\textbf {\bibinfo {volume} {128}},\
  \bibinfo {pages} {047401} (\bibinfo {year} {2022})}\BibitemShut {NoStop}%
\bibitem [{\citenamefont {Lysne}\ \emph {et~al.}(2020)\citenamefont {Lysne},
  \citenamefont {Murakami},\ and\ \citenamefont {Werner}}]{Lysne2020a}%
  \BibitemOpen
  \bibfield  {author} {\bibinfo {author} {\bibfnamefont {M.}~\bibnamefont
  {Lysne}}, \bibinfo {author} {\bibfnamefont {Y.}~\bibnamefont {Murakami}},\
  and\ \bibinfo {author} {\bibfnamefont {P.}~\bibnamefont {Werner}},\
  }\bibfield  {title} {\bibinfo {title} {{Signatures of bosonic excitations in
  high-harmonic spectra of Mott insulators}},\ }\href
  {https://doi.org/10.1103/PhysRevB.101.195139} {\bibfield  {journal} {\bibinfo
   {journal} {Phys. Rev. B}\ }\textbf {\bibinfo {volume} {101}},\ \bibinfo
  {pages} {195139} (\bibinfo {year} {2020})}\BibitemShut {NoStop}%
\bibitem [{\citenamefont {Imai}\ \emph {et~al.}(2020)\citenamefont {Imai},
  \citenamefont {Ono},\ and\ \citenamefont {Ishihara}}]{Imai2019q}%
  \BibitemOpen
  \bibfield  {author} {\bibinfo {author} {\bibfnamefont {S.}~\bibnamefont
  {Imai}}, \bibinfo {author} {\bibfnamefont {A.}~\bibnamefont {Ono}},\ and\
  \bibinfo {author} {\bibfnamefont {S.}~\bibnamefont {Ishihara}},\ }\bibfield
  {title} {\bibinfo {title} {{High Harmonic Generation in a Correlated Electron
  System}},\ }\href {https://doi.org/10.1103/PhysRevLett.124.157404} {\bibfield
   {journal} {\bibinfo  {journal} {Phys. Rev. Lett.}\ }\textbf {\bibinfo
  {volume} {124}},\ \bibinfo {pages} {157404} (\bibinfo {year}
  {2020})}\BibitemShut {NoStop}%
\bibitem [{\citenamefont {Zhu}\ \emph {et~al.}(2021)\citenamefont {Zhu},
  \citenamefont {Fauseweh}, \citenamefont {Chacon},\ and\ \citenamefont
  {Zhu}}]{Zhu2021a}%
  \BibitemOpen
  \bibfield  {author} {\bibinfo {author} {\bibfnamefont {W.}~\bibnamefont
  {Zhu}}, \bibinfo {author} {\bibfnamefont {B.}~\bibnamefont {Fauseweh}},
  \bibinfo {author} {\bibfnamefont {A.}~\bibnamefont {Chacon}},\ and\ \bibinfo
  {author} {\bibfnamefont {J.-X.}\ \bibnamefont {Zhu}},\ }\bibfield  {title}
  {\bibinfo {title} {{Ultrafast laser-driven many-body dynamics and Kondo
  coherence collapse}},\ }\href {https://doi.org/10.1103/PhysRevB.103.224305}
  {\bibfield  {journal} {\bibinfo  {journal} {Phys. Rev. B}\ }\textbf {\bibinfo
  {volume} {103}},\ \bibinfo {pages} {224305} (\bibinfo {year}
  {2021})}\BibitemShut {NoStop}%
\bibitem [{\citenamefont {Fauseweh}\ and\ \citenamefont
  {Zhu}(2020)}]{Fauseweh2020b}%
  \BibitemOpen
  \bibfield  {author} {\bibinfo {author} {\bibfnamefont {B.}~\bibnamefont
  {Fauseweh}}\ and\ \bibinfo {author} {\bibfnamefont {J.-X.}\ \bibnamefont
  {Zhu}},\ }\bibfield  {title} {\bibinfo {title} {{Laser pulse driven control
  of charge and spin order in the two-dimensional Kondo lattice}},\ }\href
  {https://doi.org/10.1103/PhysRevB.102.165128} {\bibfield  {journal} {\bibinfo
   {journal} {Phys. Rev. B}\ }\textbf {\bibinfo {volume} {102}},\ \bibinfo
  {pages} {165128} (\bibinfo {year} {2020})}\BibitemShut {NoStop}%
\bibitem [{\citenamefont {Takayoshi}\ \emph {et~al.}(2019)\citenamefont
  {Takayoshi}, \citenamefont {Murakami},\ and\ \citenamefont
  {Werner}}]{Takayoshi2019}%
  \BibitemOpen
  \bibfield  {author} {\bibinfo {author} {\bibfnamefont {S.}~\bibnamefont
  {Takayoshi}}, \bibinfo {author} {\bibfnamefont {Y.}~\bibnamefont
  {Murakami}},\ and\ \bibinfo {author} {\bibfnamefont {P.}~\bibnamefont
  {Werner}},\ }\bibfield  {title} {\bibinfo {title} {{High-harmonic generation
  in quantum spin systems}},\ }\href
  {https://doi.org/10.1103/PhysRevB.99.184303} {\bibfield  {journal} {\bibinfo
  {journal} {Phys. Rev. B}\ }\textbf {\bibinfo {volume} {99}},\ \bibinfo
  {pages} {184303} (\bibinfo {year} {2019})}\BibitemShut {NoStop}%
\bibitem [{\citenamefont {Kanega}\ \emph {et~al.}(2021)\citenamefont {Kanega},
  \citenamefont {Ikeda},\ and\ \citenamefont {Sato}}]{Kanega2021a}%
  \BibitemOpen
  \bibfield  {author} {\bibinfo {author} {\bibfnamefont {M.}~\bibnamefont
  {Kanega}}, \bibinfo {author} {\bibfnamefont {T.~N.}\ \bibnamefont {Ikeda}},\
  and\ \bibinfo {author} {\bibfnamefont {M.}~\bibnamefont {Sato}},\ }\bibfield
  {title} {\bibinfo {title} {{Linear and nonlinear optical responses in Kitaev
  spin liquids}},\ }\href {https://doi.org/10.1103/PhysRevResearch.3.L032024}
  {\bibfield  {journal} {\bibinfo  {journal} {Phys. Rev. Res.}\ }\textbf
  {\bibinfo {volume} {3}},\ \bibinfo {pages} {L032024} (\bibinfo {year}
  {2021})}\BibitemShut {NoStop}%
\bibitem [{\citenamefont {Bionta}\ \emph {et~al.}(2021)\citenamefont {Bionta},
  \citenamefont {Haddad}, \citenamefont {Leblanc}, \citenamefont {Gruson},
  \citenamefont {Lassonde}, \citenamefont {Ibrahim}, \citenamefont {Chaillou},
  \citenamefont {{\'{E}}mond}, \citenamefont {Otto}, \citenamefont
  {Jim{\'{e}}nez-Gal{\'{a}}n}, \citenamefont {Silva}, \citenamefont {Ivanov},
  \citenamefont {Siwick}, \citenamefont {Chaker},\ and\ \citenamefont
  {L{\'{e}}gar{\'{e}}}}]{Bionta2021}%
  \BibitemOpen
  \bibfield  {author} {\bibinfo {author} {\bibfnamefont {M.~R.}\ \bibnamefont
  {Bionta}}, \bibinfo {author} {\bibfnamefont {E.}~\bibnamefont {Haddad}},
  \bibinfo {author} {\bibfnamefont {A.}~\bibnamefont {Leblanc}}, \bibinfo
  {author} {\bibfnamefont {V.}~\bibnamefont {Gruson}}, \bibinfo {author}
  {\bibfnamefont {P.}~\bibnamefont {Lassonde}}, \bibinfo {author}
  {\bibfnamefont {H.}~\bibnamefont {Ibrahim}}, \bibinfo {author} {\bibfnamefont
  {J.}~\bibnamefont {Chaillou}}, \bibinfo {author} {\bibfnamefont
  {N.}~\bibnamefont {{\'{E}}mond}}, \bibinfo {author} {\bibfnamefont {M.~R.}\
  \bibnamefont {Otto}}, \bibinfo {author} {\bibfnamefont
  {{\'{A}}.}~\bibnamefont {Jim{\'{e}}nez-Gal{\'{a}}n}}, \bibinfo {author}
  {\bibfnamefont {R.~E.~F.}\ \bibnamefont {Silva}}, \bibinfo {author}
  {\bibfnamefont {M.}~\bibnamefont {Ivanov}}, \bibinfo {author} {\bibfnamefont
  {B.~J.}\ \bibnamefont {Siwick}}, \bibinfo {author} {\bibfnamefont
  {M.}~\bibnamefont {Chaker}},\ and\ \bibinfo {author} {\bibfnamefont
  {F.}~\bibnamefont {L{\'{e}}gar{\'{e}}}},\ }\bibfield  {title} {\bibinfo
  {title} {{Tracking ultrafast solid-state dynamics using high harmonic
  spectroscopy}},\ }\href {https://doi.org/10.1103/PhysRevResearch.3.023250}
  {\bibfield  {journal} {\bibinfo  {journal} {Phys. Rev. Res.}\ }\textbf
  {\bibinfo {volume} {3}},\ \bibinfo {pages} {023250} (\bibinfo {year}
  {2021})}\BibitemShut {NoStop}%
\bibitem [{\citenamefont {Uchida}\ \emph {et~al.}(2022)\citenamefont {Uchida},
  \citenamefont {Mattoni}, \citenamefont {Yonezawa}, \citenamefont {Nakamura},
  \citenamefont {Maeno},\ and\ \citenamefont {Tanaka}}]{Uchida2021b}%
  \BibitemOpen
  \bibfield  {author} {\bibinfo {author} {\bibfnamefont {K.}~\bibnamefont
  {Uchida}}, \bibinfo {author} {\bibfnamefont {G.}~\bibnamefont {Mattoni}},
  \bibinfo {author} {\bibfnamefont {S.}~\bibnamefont {Yonezawa}}, \bibinfo
  {author} {\bibfnamefont {F.}~\bibnamefont {Nakamura}}, \bibinfo {author}
  {\bibfnamefont {Y.}~\bibnamefont {Maeno}},\ and\ \bibinfo {author}
  {\bibfnamefont {K.}~\bibnamefont {Tanaka}},\ }\bibfield  {title} {\bibinfo
  {title} {High-order harmonic generation and its unconventional scaling law in
  the mott-insulating ${\mathrm{ca}}_{2}{\mathrm{ruo}}_{4}$},\ }\href
  {https://doi.org/10.1103/PhysRevLett.128.127401} {\bibfield  {journal}
  {\bibinfo  {journal} {Phys. Rev. Lett.}\ }\textbf {\bibinfo {volume} {128}},\
  \bibinfo {pages} {127401} (\bibinfo {year} {2022})}\BibitemShut {NoStop}%
\bibitem [{\citenamefont {Murakami}\ \emph {et~al.}(2022)\citenamefont
  {Murakami}, \citenamefont {Uchida}, \citenamefont {Koga}, \citenamefont
  {Tanaka},\ and\ \citenamefont {Werner}}]{Murakami2022b}%
  \BibitemOpen
  \bibfield  {author} {\bibinfo {author} {\bibfnamefont {Y.}~\bibnamefont
  {Murakami}}, \bibinfo {author} {\bibfnamefont {K.}~\bibnamefont {Uchida}},
  \bibinfo {author} {\bibfnamefont {A.}~\bibnamefont {Koga}}, \bibinfo {author}
  {\bibfnamefont {K.}~\bibnamefont {Tanaka}},\ and\ \bibinfo {author}
  {\bibfnamefont {P.}~\bibnamefont {Werner}},\ }\bibfield  {title} {\bibinfo
  {title} {{Anomalous Temperature Dependence of High-Harmonic Generation in
  Mott Insulators}},\ }\href {https://doi.org/10.1103/PhysRevLett.129.157401}
  {\bibfield  {journal} {\bibinfo  {journal} {Phys. Rev. Lett.}\ }\textbf
  {\bibinfo {volume} {129}},\ \bibinfo {pages} {157401} (\bibinfo {year}
  {2022})}\BibitemShut {NoStop}%
\bibitem [{\citenamefont {Gr{\aa}n{\"{a}}s}\ \emph {et~al.}(2022)\citenamefont
  {Gr{\aa}n{\"{a}}s}, \citenamefont {Vaskivskyi}, \citenamefont {Wang},
  \citenamefont {Thunstr{\"{o}}m}, \citenamefont {Ghimire}, \citenamefont
  {Knut}, \citenamefont {S{\"{o}}derstr{\"{o}}m}, \citenamefont {Kjellsson},
  \citenamefont {Turenne}, \citenamefont {Engel}, \citenamefont {Beye},
  \citenamefont {Lu}, \citenamefont {Higley}, \citenamefont {Reid},
  \citenamefont {Schlotter}, \citenamefont {Coslovich}, \citenamefont
  {Hoffmann}, \citenamefont {Kolesov}, \citenamefont
  {Sch{\"{u}}{\ss}ler-Langeheine}, \citenamefont {Styervoyedov}, \citenamefont
  {Tancogne-Dejean}, \citenamefont {Sentef}, \citenamefont {Reis},
  \citenamefont {Rubio}, \citenamefont {Parkin}, \citenamefont {Karis},
  \citenamefont {Rubensson}, \citenamefont {Eriksson},\ and\ \citenamefont
  {D{\"{u}}rr}}]{Granas2022}%
  \BibitemOpen
  \bibfield  {author} {\bibinfo {author} {\bibfnamefont {O.}~\bibnamefont
  {Gr{\aa}n{\"{a}}s}}, \bibinfo {author} {\bibfnamefont {I.}~\bibnamefont
  {Vaskivskyi}}, \bibinfo {author} {\bibfnamefont {X.}~\bibnamefont {Wang}},
  \bibinfo {author} {\bibfnamefont {P.}~\bibnamefont {Thunstr{\"{o}}m}},
  \bibinfo {author} {\bibfnamefont {S.}~\bibnamefont {Ghimire}}, \bibinfo
  {author} {\bibfnamefont {R.}~\bibnamefont {Knut}}, \bibinfo {author}
  {\bibfnamefont {J.}~\bibnamefont {S{\"{o}}derstr{\"{o}}m}}, \bibinfo {author}
  {\bibfnamefont {L.}~\bibnamefont {Kjellsson}}, \bibinfo {author}
  {\bibfnamefont {D.}~\bibnamefont {Turenne}}, \bibinfo {author} {\bibfnamefont
  {R.~Y.}\ \bibnamefont {Engel}}, \bibinfo {author} {\bibfnamefont
  {M.}~\bibnamefont {Beye}}, \bibinfo {author} {\bibfnamefont {J.}~\bibnamefont
  {Lu}}, \bibinfo {author} {\bibfnamefont {D.~J.}\ \bibnamefont {Higley}},
  \bibinfo {author} {\bibfnamefont {A.~H.}\ \bibnamefont {Reid}}, \bibinfo
  {author} {\bibfnamefont {W.}~\bibnamefont {Schlotter}}, \bibinfo {author}
  {\bibfnamefont {G.}~\bibnamefont {Coslovich}}, \bibinfo {author}
  {\bibfnamefont {M.}~\bibnamefont {Hoffmann}}, \bibinfo {author}
  {\bibfnamefont {G.}~\bibnamefont {Kolesov}}, \bibinfo {author} {\bibfnamefont
  {C.}~\bibnamefont {Sch{\"{u}}{\ss}ler-Langeheine}}, \bibinfo {author}
  {\bibfnamefont {A.}~\bibnamefont {Styervoyedov}}, \bibinfo {author}
  {\bibfnamefont {N.}~\bibnamefont {Tancogne-Dejean}}, \bibinfo {author}
  {\bibfnamefont {M.~A.}\ \bibnamefont {Sentef}}, \bibinfo {author}
  {\bibfnamefont {D.~A.}\ \bibnamefont {Reis}}, \bibinfo {author}
  {\bibfnamefont {A.}~\bibnamefont {Rubio}}, \bibinfo {author} {\bibfnamefont
  {S.~S.~P.}\ \bibnamefont {Parkin}}, \bibinfo {author} {\bibfnamefont
  {O.}~\bibnamefont {Karis}}, \bibinfo {author} {\bibfnamefont {J.-E.}\
  \bibnamefont {Rubensson}}, \bibinfo {author} {\bibfnamefont {O.}~\bibnamefont
  {Eriksson}},\ and\ \bibinfo {author} {\bibfnamefont {H.~A.}\ \bibnamefont
  {D{\"{u}}rr}},\ }\bibfield  {title} {\bibinfo {title} {{Ultrafast
  modification of the electronic structure of a correlated insulator}},\ }\href
  {https://doi.org/10.1103/PhysRevResearch.4.L032030} {\bibfield  {journal}
  {\bibinfo  {journal} {Phys. Rev. Res.}\ }\textbf {\bibinfo {volume} {4}},\
  \bibinfo {pages} {L032030} (\bibinfo {year} {2022})}\BibitemShut {NoStop}%
\bibitem [{\citenamefont {Alcal{\`{a}}}\ \emph {et~al.}(2022)\citenamefont
  {Alcal{\`{a}}}, \citenamefont {Bhattacharya}, \citenamefont {Biegert},
  \citenamefont {Ciappina}, \citenamefont {Elu}, \citenamefont {Gra{\ss}},
  \citenamefont {Grochowski}, \citenamefont {Lewenstein}, \citenamefont
  {Palau}, \citenamefont {Sidiropoulos}, \citenamefont {Steinle},\ and\
  \citenamefont {Tyulnev}}]{Alcala2022a}%
  \BibitemOpen
  \bibfield  {author} {\bibinfo {author} {\bibfnamefont {J.}~\bibnamefont
  {Alcal{\`{a}}}}, \bibinfo {author} {\bibfnamefont {U.}~\bibnamefont
  {Bhattacharya}}, \bibinfo {author} {\bibfnamefont {J.}~\bibnamefont
  {Biegert}}, \bibinfo {author} {\bibfnamefont {M.}~\bibnamefont {Ciappina}},
  \bibinfo {author} {\bibfnamefont {U.}~\bibnamefont {Elu}}, \bibinfo {author}
  {\bibfnamefont {T.}~\bibnamefont {Gra{\ss}}}, \bibinfo {author}
  {\bibfnamefont {P.~T.}\ \bibnamefont {Grochowski}}, \bibinfo {author}
  {\bibfnamefont {M.}~\bibnamefont {Lewenstein}}, \bibinfo {author}
  {\bibfnamefont {A.}~\bibnamefont {Palau}}, \bibinfo {author} {\bibfnamefont
  {T.~P.~H.}\ \bibnamefont {Sidiropoulos}}, \bibinfo {author} {\bibfnamefont
  {T.}~\bibnamefont {Steinle}},\ and\ \bibinfo {author} {\bibfnamefont
  {I.}~\bibnamefont {Tyulnev}},\ }\bibfield  {title} {\bibinfo {title}
  {{High-harmonic spectroscopy of quantum phase transitions in a high-Tc
  superconductor}},\ }\href {https://doi.org/10.1073/pnas.2207766119}
  {\bibfield  {journal} {\bibinfo  {journal} {Proc. Natl. Acad. Sci.}\ }\textbf
  {\bibinfo {volume} {119}},\ \bibinfo {pages} {1} (\bibinfo {year}
  {2022})}\BibitemShut {NoStop}%
\bibitem [{\citenamefont {Garg}\ \emph
  {et~al.}(2016{\natexlab{b}})\citenamefont {Garg}, \citenamefont {Zhan},
  \citenamefont {Luu}, \citenamefont {Lakhotia}, \citenamefont {Klostermann},
  \citenamefont {Guggenmos},\ and\ \citenamefont {Goulielmakis}}]{Garg2016}%
  \BibitemOpen
  \bibfield  {author} {\bibinfo {author} {\bibfnamefont {M.}~\bibnamefont
  {Garg}}, \bibinfo {author} {\bibfnamefont {M.}~\bibnamefont {Zhan}}, \bibinfo
  {author} {\bibfnamefont {T.~T.}\ \bibnamefont {Luu}}, \bibinfo {author}
  {\bibfnamefont {H.}~\bibnamefont {Lakhotia}}, \bibinfo {author}
  {\bibfnamefont {T.}~\bibnamefont {Klostermann}}, \bibinfo {author}
  {\bibfnamefont {A.}~\bibnamefont {Guggenmos}},\ and\ \bibinfo {author}
  {\bibfnamefont {E.}~\bibnamefont {Goulielmakis}},\ }\bibfield  {title}
  {\bibinfo {title} {{Multi-petahertz electronic metrology}},\ }\href
  {https://doi.org/10.1038/nature19821} {\bibfield  {journal} {\bibinfo
  {journal} {Nature}\ }\textbf {\bibinfo {volume} {538}},\ \bibinfo {pages}
  {359} (\bibinfo {year} {2016}{\natexlab{b}})}\BibitemShut {NoStop}%
\bibitem [{\citenamefont {Garg}\ \emph {et~al.}(2018)\citenamefont {Garg},
  \citenamefont {Kim},\ and\ \citenamefont {Goulielmakis}}]{Garg2018}%
  \BibitemOpen
  \bibfield  {author} {\bibinfo {author} {\bibfnamefont {M.}~\bibnamefont
  {Garg}}, \bibinfo {author} {\bibfnamefont {H.~Y.}\ \bibnamefont {Kim}},\ and\
  \bibinfo {author} {\bibfnamefont {E.}~\bibnamefont {Goulielmakis}},\
  }\bibfield  {title} {\bibinfo {title} {{Ultimate waveform reproducibility of
  extreme-ultraviolet pulses by high-harmonic generation in quartz}},\ }\href
  {https://doi.org/10.1038/s41566-018-0123-6} {\bibfield  {journal} {\bibinfo
  {journal} {Nat. Photonics}\ }\textbf {\bibinfo {volume} {12}},\ \bibinfo
  {pages} {291} (\bibinfo {year} {2018})}\BibitemShut {NoStop}%
\bibitem [{\citenamefont {Guan}\ \emph {et~al.}(2016)\citenamefont {Guan},
  \citenamefont {Zhou},\ and\ \citenamefont {Bian}}]{Guan2016}%
  \BibitemOpen
  \bibfield  {author} {\bibinfo {author} {\bibfnamefont {Z.}~\bibnamefont
  {Guan}}, \bibinfo {author} {\bibfnamefont {X.-X.}\ \bibnamefont {Zhou}},\
  and\ \bibinfo {author} {\bibfnamefont {X.-B.}\ \bibnamefont {Bian}},\
  }\bibfield  {title} {\bibinfo {title} {{High-order-harmonic generation from
  periodic potentials driven by few-cycle laser pulses}},\ }\href
  {https://doi.org/10.1103/PhysRevA.93.033852} {\bibfield  {journal} {\bibinfo
  {journal} {Phys. Rev. A}\ }\textbf {\bibinfo {volume} {93}},\ \bibinfo
  {pages} {033852} (\bibinfo {year} {2016})}\BibitemShut {NoStop}%
\bibitem [{\citenamefont {Song}\ \emph {et~al.}(2019)\citenamefont {Song},
  \citenamefont {Zuo}, \citenamefont {Yang}, \citenamefont {Li}, \citenamefont
  {Meier},\ and\ \citenamefont {Yang}}]{Song2019}%
  \BibitemOpen
  \bibfield  {author} {\bibinfo {author} {\bibfnamefont {X.}~\bibnamefont
  {Song}}, \bibinfo {author} {\bibfnamefont {R.}~\bibnamefont {Zuo}}, \bibinfo
  {author} {\bibfnamefont {S.}~\bibnamefont {Yang}}, \bibinfo {author}
  {\bibfnamefont {P.}~\bibnamefont {Li}}, \bibinfo {author} {\bibfnamefont
  {T.}~\bibnamefont {Meier}},\ and\ \bibinfo {author} {\bibfnamefont
  {W.}~\bibnamefont {Yang}},\ }\bibfield  {title} {\bibinfo {title}
  {{Attosecond temporal confinement of interband excitation by intraband
  motion}},\ }\href {https://doi.org/10.1364/OE.27.002225} {\bibfield
  {journal} {\bibinfo  {journal} {Opt. Express}\ }\textbf {\bibinfo {volume}
  {27}},\ \bibinfo {pages} {2225} (\bibinfo {year} {2019})}\BibitemShut
  {NoStop}%
\bibitem [{\citenamefont {Nourbakhsh}\ \emph {et~al.}(2021)\citenamefont
  {Nourbakhsh}, \citenamefont {Tancogne-Dejean}, \citenamefont {Merdji},\ and\
  \citenamefont {Rubio}}]{Nourbakhsh2021}%
  \BibitemOpen
  \bibfield  {author} {\bibinfo {author} {\bibfnamefont {Z.}~\bibnamefont
  {Nourbakhsh}}, \bibinfo {author} {\bibfnamefont {N.}~\bibnamefont
  {Tancogne-Dejean}}, \bibinfo {author} {\bibfnamefont {H.}~\bibnamefont
  {Merdji}},\ and\ \bibinfo {author} {\bibfnamefont {A.}~\bibnamefont
  {Rubio}},\ }\bibfield  {title} {\bibinfo {title} {{High Harmonics and
  Isolated Attosecond Pulses from MgO}},\ }\href
  {https://doi.org/10.1103/PhysRevApplied.15.014013} {\bibfield  {journal}
  {\bibinfo  {journal} {Phys. Rev. Appl.}\ }\textbf {\bibinfo {volume} {15}},\
  \bibinfo {pages} {014013} (\bibinfo {year} {2021})}\BibitemShut {NoStop}%
\bibitem [{\citenamefont {Sadeghifaraz}\ \emph {et~al.}(2022)\citenamefont
  {Sadeghifaraz}, \citenamefont {Irani},\ and\ \citenamefont
  {Monfared}}]{Sadeghifaraz2022}%
  \BibitemOpen
  \bibfield  {author} {\bibinfo {author} {\bibfnamefont {A.}~\bibnamefont
  {Sadeghifaraz}}, \bibinfo {author} {\bibfnamefont {E.}~\bibnamefont
  {Irani}},\ and\ \bibinfo {author} {\bibfnamefont {M.}~\bibnamefont
  {Monfared}},\ }\bibfield  {title} {\bibinfo {title} {{Efficient attosecond
  pulse generation from WS2 semiconductor by tailoring the driving laser
  pulse}},\ }\href {https://doi.org/10.1016/j.optcom.2022.128226} {\bibfield
  {journal} {\bibinfo  {journal} {Opt. Commun.}\ }\textbf {\bibinfo {volume}
  {516}},\ \bibinfo {pages} {128226} (\bibinfo {year} {2022})}\BibitemShut
  {NoStop}%
\bibitem [{\citenamefont {Hassan}\ \emph {et~al.}(2012)\citenamefont {Hassan},
  \citenamefont {Wirth}, \citenamefont {Grgura{\v{s}}}, \citenamefont {Moulet},
  \citenamefont {Luu}, \citenamefont {Gagnon}, \citenamefont {Pervak},\ and\
  \citenamefont {Goulielmakis}}]{Hassan2012}%
  \BibitemOpen
  \bibfield  {author} {\bibinfo {author} {\bibfnamefont {M.~T.}\ \bibnamefont
  {Hassan}}, \bibinfo {author} {\bibfnamefont {A.}~\bibnamefont {Wirth}},
  \bibinfo {author} {\bibfnamefont {I.}~\bibnamefont {Grgura{\v{s}}}}, \bibinfo
  {author} {\bibfnamefont {A.}~\bibnamefont {Moulet}}, \bibinfo {author}
  {\bibfnamefont {T.~T.}\ \bibnamefont {Luu}}, \bibinfo {author} {\bibfnamefont
  {J.}~\bibnamefont {Gagnon}}, \bibinfo {author} {\bibfnamefont
  {V.}~\bibnamefont {Pervak}},\ and\ \bibinfo {author} {\bibfnamefont
  {E.}~\bibnamefont {Goulielmakis}},\ }\bibfield  {title} {\bibinfo {title}
  {{Invited Article: Attosecond photonics: Synthesis and control of light
  transients}},\ }\href {https://doi.org/10.1063/1.4758310} {\bibfield
  {journal} {\bibinfo  {journal} {Rev. Sci. Instrum.}\ }\textbf {\bibinfo
  {volume} {83}},\ \bibinfo {pages} {111301} (\bibinfo {year}
  {2012})}\BibitemShut {NoStop}%
\bibitem [{\citenamefont {Mucke}\ \emph {et~al.}(2015)\citenamefont {Mucke},
  \citenamefont {Fang}, \citenamefont {Cirmi}, \citenamefont {Rossi},
  \citenamefont {Chia}, \citenamefont {Ye}, \citenamefont {Yang}, \citenamefont
  {Mainz}, \citenamefont {Manzoni}, \citenamefont {Farinello}, \citenamefont
  {Cerullo},\ and\ \citenamefont {Kartner}}]{Mucke2015}%
  \BibitemOpen
  \bibfield  {author} {\bibinfo {author} {\bibfnamefont {O.~D.}\ \bibnamefont
  {Mucke}}, \bibinfo {author} {\bibfnamefont {S.}~\bibnamefont {Fang}},
  \bibinfo {author} {\bibfnamefont {G.}~\bibnamefont {Cirmi}}, \bibinfo
  {author} {\bibfnamefont {G.~M.}\ \bibnamefont {Rossi}}, \bibinfo {author}
  {\bibfnamefont {S.-H.}\ \bibnamefont {Chia}}, \bibinfo {author}
  {\bibfnamefont {H.}~\bibnamefont {Ye}}, \bibinfo {author} {\bibfnamefont
  {Y.}~\bibnamefont {Yang}}, \bibinfo {author} {\bibfnamefont {R.}~\bibnamefont
  {Mainz}}, \bibinfo {author} {\bibfnamefont {C.}~\bibnamefont {Manzoni}},
  \bibinfo {author} {\bibfnamefont {P.}~\bibnamefont {Farinello}}, \bibinfo
  {author} {\bibfnamefont {G.}~\bibnamefont {Cerullo}},\ and\ \bibinfo {author}
  {\bibfnamefont {F.~X.}\ \bibnamefont {Kartner}},\ }\bibfield  {title}
  {\bibinfo {title} {{Toward Waveform Nonlinear Optics Using Multimillijoule
  Sub-Cycle Waveform Synthesizers}},\ }\href
  {https://doi.org/10.1109/JSTQE.2015.2426653} {\bibfield  {journal} {\bibinfo
  {journal} {IEEE J. Sel. Top. Quantum Electron.}\ }\textbf {\bibinfo {volume}
  {21}},\ \bibinfo {pages} {1} (\bibinfo {year} {2015})}\BibitemShut {NoStop}%
\bibitem [{\citenamefont {Lin}\ \emph {et~al.}(2020)\citenamefont {Lin},
  \citenamefont {Nabekawa},\ and\ \citenamefont {Midorikawa}}]{Lin2020}%
  \BibitemOpen
  \bibfield  {author} {\bibinfo {author} {\bibfnamefont {Y.-C.}\ \bibnamefont
  {Lin}}, \bibinfo {author} {\bibfnamefont {Y.}~\bibnamefont {Nabekawa}},\ and\
  \bibinfo {author} {\bibfnamefont {K.}~\bibnamefont {Midorikawa}},\ }\bibfield
   {title} {\bibinfo {title} {{Optical parametric amplification of sub-cycle
  shortwave infrared pulses}},\ }\href
  {https://doi.org/10.1038/s41467-020-17247-9} {\bibfield  {journal} {\bibinfo
  {journal} {Nat. Commun.}\ }\textbf {\bibinfo {volume} {11}},\ \bibinfo
  {pages} {3413} (\bibinfo {year} {2020})}\BibitemShut {NoStop}%
\bibitem [{\citenamefont {Tian}\ \emph {et~al.}(2021)\citenamefont {Tian},
  \citenamefont {He}, \citenamefont {Yang},\ and\ \citenamefont
  {Liang}}]{Tian2021}%
  \BibitemOpen
  \bibfield  {author} {\bibinfo {author} {\bibfnamefont {K.}~\bibnamefont
  {Tian}}, \bibinfo {author} {\bibfnamefont {L.}~\bibnamefont {He}}, \bibinfo
  {author} {\bibfnamefont {X.}~\bibnamefont {Yang}},\ and\ \bibinfo {author}
  {\bibfnamefont {H.}~\bibnamefont {Liang}},\ }\bibfield  {title} {\bibinfo
  {title} {{Mid-Infrared Few-Cycle Pulse Generation and Amplification}},\
  }\href {https://doi.org/10.3390/photonics8080290} {\bibfield  {journal}
  {\bibinfo  {journal} {Photonics}\ }\textbf {\bibinfo {volume} {8}},\ \bibinfo
  {pages} {290} (\bibinfo {year} {2021})}\BibitemShut {NoStop}%
\bibitem [{\citenamefont {Alqattan}\ \emph {et~al.}(2022)\citenamefont
  {Alqattan}, \citenamefont {Hui}, \citenamefont {Pervak},\ and\ \citenamefont
  {Hassan}}]{Alqattan2022}%
  \BibitemOpen
  \bibfield  {author} {\bibinfo {author} {\bibfnamefont {H.}~\bibnamefont
  {Alqattan}}, \bibinfo {author} {\bibfnamefont {D.}~\bibnamefont {Hui}},
  \bibinfo {author} {\bibfnamefont {V.}~\bibnamefont {Pervak}},\ and\ \bibinfo
  {author} {\bibfnamefont {M.~T.}\ \bibnamefont {Hassan}},\ }\bibfield  {title}
  {\bibinfo {title} {{Attosecond light field synthesis}},\ }\href
  {https://doi.org/10.1063/5.0082958} {\bibfield  {journal} {\bibinfo
  {journal} {APL Photonics}\ }\textbf {\bibinfo {volume} {7}},\ \bibinfo
  {pages} {041301} (\bibinfo {year} {2022})}\BibitemShut {NoStop}%
\bibitem [{\citenamefont {Su}\ \emph {et~al.}(2022)\citenamefont {Su},
  \citenamefont {Fang}, \citenamefont {Wang}, \citenamefont {Liang},
  \citenamefont {Chang}, \citenamefont {He},\ and\ \citenamefont
  {Wei}}]{Su2022}%
  \BibitemOpen
  \bibfield  {author} {\bibinfo {author} {\bibfnamefont {Y.}~\bibnamefont
  {Su}}, \bibinfo {author} {\bibfnamefont {S.}~\bibnamefont {Fang}}, \bibinfo
  {author} {\bibfnamefont {S.}~\bibnamefont {Wang}}, \bibinfo {author}
  {\bibfnamefont {Y.}~\bibnamefont {Liang}}, \bibinfo {author} {\bibfnamefont
  {G.}~\bibnamefont {Chang}}, \bibinfo {author} {\bibfnamefont
  {X.}~\bibnamefont {He}},\ and\ \bibinfo {author} {\bibfnamefont
  {Z.}~\bibnamefont {Wei}},\ }\bibfield  {title} {\bibinfo {title} {{Optimal
  generation of delay-controlled few-cycle pulses for high harmonic generation
  in solids}},\ }\href {https://doi.org/10.1063/5.0085472} {\bibfield
  {journal} {\bibinfo  {journal} {Appl. Phys. Lett.}\ }\textbf {\bibinfo
  {volume} {120}},\ \bibinfo {pages} {121105} (\bibinfo {year}
  {2022})}\BibitemShut {NoStop}%
\bibitem [{\citenamefont {Gaumnitz}\ \emph {et~al.}(2017)\citenamefont
  {Gaumnitz}, \citenamefont {Jain}, \citenamefont {Pertot}, \citenamefont
  {Huppert}, \citenamefont {Jordan}, \citenamefont {Ardana-Lamas},\ and\
  \citenamefont {W{\"{o}}rner}}]{Gaumnitz2017}%
  \BibitemOpen
  \bibfield  {author} {\bibinfo {author} {\bibfnamefont {T.}~\bibnamefont
  {Gaumnitz}}, \bibinfo {author} {\bibfnamefont {A.}~\bibnamefont {Jain}},
  \bibinfo {author} {\bibfnamefont {Y.}~\bibnamefont {Pertot}}, \bibinfo
  {author} {\bibfnamefont {M.}~\bibnamefont {Huppert}}, \bibinfo {author}
  {\bibfnamefont {I.}~\bibnamefont {Jordan}}, \bibinfo {author} {\bibfnamefont
  {F.}~\bibnamefont {Ardana-Lamas}},\ and\ \bibinfo {author} {\bibfnamefont
  {H.~J.}\ \bibnamefont {W{\"{o}}rner}},\ }\bibfield  {title} {\bibinfo {title}
  {{Streaking of 43-attosecond soft-X-ray pulses generated by a passively
  CEP-stable mid-infrared driver}},\ }\href
  {https://doi.org/10.1364/OE.25.027506} {\bibfield  {journal} {\bibinfo
  {journal} {Opt. Express}\ }\textbf {\bibinfo {volume} {25}},\ \bibinfo
  {pages} {27506} (\bibinfo {year} {2017})}\BibitemShut {NoStop}%
\bibitem [{\citenamefont {Shafir}\ \emph {et~al.}(2012)\citenamefont {Shafir},
  \citenamefont {Soifer}, \citenamefont {Bruner}, \citenamefont {Dagan},
  \citenamefont {Mairesse}, \citenamefont {Patchkovskii}, \citenamefont
  {Ivanov}, \citenamefont {Smirnova},\ and\ \citenamefont
  {Dudovich}}]{Shafir2012}%
  \BibitemOpen
  \bibfield  {author} {\bibinfo {author} {\bibfnamefont {D.}~\bibnamefont
  {Shafir}}, \bibinfo {author} {\bibfnamefont {H.}~\bibnamefont {Soifer}},
  \bibinfo {author} {\bibfnamefont {B.~D.}\ \bibnamefont {Bruner}}, \bibinfo
  {author} {\bibfnamefont {M.}~\bibnamefont {Dagan}}, \bibinfo {author}
  {\bibfnamefont {Y.}~\bibnamefont {Mairesse}}, \bibinfo {author}
  {\bibfnamefont {S.}~\bibnamefont {Patchkovskii}}, \bibinfo {author}
  {\bibfnamefont {M.~Y.}\ \bibnamefont {Ivanov}}, \bibinfo {author}
  {\bibfnamefont {O.}~\bibnamefont {Smirnova}},\ and\ \bibinfo {author}
  {\bibfnamefont {N.}~\bibnamefont {Dudovich}},\ }\bibfield  {title} {\bibinfo
  {title} {{Resolving the time when an electron exits a tunnelling barrier}},\
  }\href {https://doi.org/10.1038/nature11025} {\bibfield  {journal} {\bibinfo
  {journal} {Nature}\ }\textbf {\bibinfo {volume} {485}},\ \bibinfo {pages}
  {343} (\bibinfo {year} {2012})}\BibitemShut {NoStop}%
\bibitem [{\citenamefont {Vitanov}\ and\ \citenamefont
  {Garraway}(1996)}]{Vitanov1996}%
  \BibitemOpen
  \bibfield  {author} {\bibinfo {author} {\bibfnamefont {N.~V.}\ \bibnamefont
  {Vitanov}}\ and\ \bibinfo {author} {\bibfnamefont {B.~M.}\ \bibnamefont
  {Garraway}},\ }\bibfield  {title} {\bibinfo {title} {{Landau-Zener model:
  Effects of finite coupling duration}},\ }\href
  {https://doi.org/10.1103/PhysRevA.53.4288} {\bibfield  {journal} {\bibinfo
  {journal} {Phys. Rev. A}\ }\textbf {\bibinfo {volume} {53}},\ \bibinfo
  {pages} {4288} (\bibinfo {year} {1996})}\BibitemShut {NoStop}%
\bibitem [{\citenamefont {Oka}(2012)}]{Oka2012a}%
  \BibitemOpen
  \bibfield  {author} {\bibinfo {author} {\bibfnamefont {T.}~\bibnamefont
  {Oka}},\ }\bibfield  {title} {\bibinfo {title} {{Nonlinear doublon production
  in a Mott insulator: Landau-Dykhne method applied to an integrable model}},\
  }\href {https://doi.org/10.1103/PhysRevB.86.075148} {\bibfield  {journal}
  {\bibinfo  {journal} {Phys. Rev. B}\ }\textbf {\bibinfo {volume} {86}},\
  \bibinfo {pages} {075148} (\bibinfo {year} {2012})}\BibitemShut {NoStop}%
\bibitem [{\citenamefont {MacColl}(1932)}]{MacColl1932}%
  \BibitemOpen
  \bibfield  {author} {\bibinfo {author} {\bibfnamefont {L.~A.}\ \bibnamefont
  {MacColl}},\ }\bibfield  {title} {\bibinfo {title} {{Note on the Transmission
  and Reflection of Wave Packets by Potential Barriers}},\ }\href
  {https://doi.org/10.1103/PhysRev.40.621} {\bibfield  {journal} {\bibinfo
  {journal} {Phys. Rev.}\ }\textbf {\bibinfo {volume} {40}},\ \bibinfo {pages}
  {621} (\bibinfo {year} {1932})}\BibitemShut {NoStop}%
\bibitem [{\citenamefont {Yakaboylu}\ \emph {et~al.}(2014)\citenamefont
  {Yakaboylu}, \citenamefont {Klaiber},\ and\ \citenamefont
  {Hatsagortsyan}}]{Yakaboylu2014}%
  \BibitemOpen
  \bibfield  {author} {\bibinfo {author} {\bibfnamefont {E.}~\bibnamefont
  {Yakaboylu}}, \bibinfo {author} {\bibfnamefont {M.}~\bibnamefont {Klaiber}},\
  and\ \bibinfo {author} {\bibfnamefont {K.~Z.}\ \bibnamefont
  {Hatsagortsyan}},\ }\bibfield  {title} {\bibinfo {title} {{Wigner time delay
  for tunneling ionization via the electron propagator}},\ }\href
  {https://doi.org/10.1103/PhysRevA.90.012116} {\bibfield  {journal} {\bibinfo
  {journal} {Phys. Rev. A}\ }\textbf {\bibinfo {volume} {90}},\ \bibinfo
  {pages} {012116} (\bibinfo {year} {2014})}\BibitemShut {NoStop}%
\bibitem [{\citenamefont {Kheifets}(2020)}]{Kheifets2020}%
  \BibitemOpen
  \bibfield  {author} {\bibinfo {author} {\bibfnamefont {A.~S.}\ \bibnamefont
  {Kheifets}},\ }\bibfield  {title} {\bibinfo {title} {{The attoclock and the
  tunneling time debate}},\ }\href {https://doi.org/10.1088/1361-6455/ab6b3b}
  {\bibfield  {journal} {\bibinfo  {journal} {J. Phys. B}\ }\textbf {\bibinfo
  {volume} {53}},\ \bibinfo {pages} {072001} (\bibinfo {year}
  {2020})}\BibitemShut {NoStop}%
\bibitem [{\citenamefont {Imai}\ \emph {et~al.}(2022)\citenamefont {Imai},
  \citenamefont {Ono},\ and\ \citenamefont {Ishihara}}]{Imai2022a}%
  \BibitemOpen
  \bibfield  {author} {\bibinfo {author} {\bibfnamefont {S.}~\bibnamefont
  {Imai}}, \bibinfo {author} {\bibfnamefont {A.}~\bibnamefont {Ono}},\ and\
  \bibinfo {author} {\bibfnamefont {S.}~\bibnamefont {Ishihara}},\ }\bibfield
  {title} {\bibinfo {title} {{Energy-band echoes: Time-reversed light emission
  from optically driven quasiparticle wave packets}},\ }\href
  {https://doi.org/10.1103/PhysRevResearch.4.043155} {\bibfield  {journal}
  {\bibinfo  {journal} {Phys. Rev. Res.}\ }\textbf {\bibinfo {volume} {4}},\
  \bibinfo {pages} {043155} (\bibinfo {year} {2022})}\BibitemShut {NoStop}%
\bibitem [{\citenamefont {Narozhnyi}\ and\ \citenamefont
  {Nikishov}(1970)}]{osti_4159359}%
  \BibitemOpen
  \bibfield  {author} {\bibinfo {author} {\bibfnamefont {N.~B.}\ \bibnamefont
  {Narozhnyi}}\ and\ \bibinfo {author} {\bibfnamefont {A.~I.}\ \bibnamefont
  {Nikishov}},\ }\bibfield  {title} {\bibinfo {title} {The simplest processes
  in the pair-creating electric field},\ }\href
  {https://www.osti.gov/biblio/4159359} {\bibfield  {journal} {\bibinfo
  {journal} {Sov. J. Nucl. Phys.}\ }\textbf {\bibinfo {volume} {11}},\ \bibinfo
  {pages} {596} (\bibinfo {year} {1970})}\BibitemShut {NoStop}%
\bibitem [{Note1()}]{Note1}%
  \BibitemOpen
  \bibinfo {note} {The minimum FWHM ($\approx 625~\protect \mathrm {as}$) is
  located away from $\updelta A_{\protect \mathrm {d}} = 0$, because we used
  the same $\tau _{\protect \mathrm {r}}$ for $|\updelta A_{\protect \mathrm
  {d}}| \neq 0$.}\BibitemShut {Stop}%
\bibitem [{\citenamefont {Geissler}\ \emph {et~al.}(1999)\citenamefont
  {Geissler}, \citenamefont {Tempea}, \citenamefont {Scrinzi}, \citenamefont
  {Schn{\"{u}}rer}, \citenamefont {Krausz},\ and\ \citenamefont
  {Brabec}}]{Geissler1999}%
  \BibitemOpen
  \bibfield  {author} {\bibinfo {author} {\bibfnamefont {M.}~\bibnamefont
  {Geissler}}, \bibinfo {author} {\bibfnamefont {G.}~\bibnamefont {Tempea}},
  \bibinfo {author} {\bibfnamefont {A.}~\bibnamefont {Scrinzi}}, \bibinfo
  {author} {\bibfnamefont {M.}~\bibnamefont {Schn{\"{u}}rer}}, \bibinfo
  {author} {\bibfnamefont {F.}~\bibnamefont {Krausz}},\ and\ \bibinfo {author}
  {\bibfnamefont {T.}~\bibnamefont {Brabec}},\ }\bibfield  {title} {\bibinfo
  {title} {{Light Propagation in Field-Ionizing Media: Extreme Nonlinear
  Optics}},\ }\href {https://doi.org/10.1103/PhysRevLett.83.2930} {\bibfield
  {journal} {\bibinfo  {journal} {Phys. Rev. Lett.}\ }\textbf {\bibinfo
  {volume} {83}},\ \bibinfo {pages} {2930} (\bibinfo {year}
  {1999})}\BibitemShut {NoStop}%
\bibitem [{\citenamefont {J{\"{u}}rgens}\ \emph {et~al.}(2020)\citenamefont
  {J{\"{u}}rgens}, \citenamefont {Liewehr}, \citenamefont {Kruse},
  \citenamefont {Peltz}, \citenamefont {Engel}, \citenamefont {Husakou},
  \citenamefont {Witting}, \citenamefont {Ivanov}, \citenamefont {Vrakking},
  \citenamefont {Fennel},\ and\ \citenamefont
  {Mermillod-Blondin}}]{Jurgens2020}%
  \BibitemOpen
  \bibfield  {author} {\bibinfo {author} {\bibfnamefont {P.}~\bibnamefont
  {J{\"{u}}rgens}}, \bibinfo {author} {\bibfnamefont {B.}~\bibnamefont
  {Liewehr}}, \bibinfo {author} {\bibfnamefont {B.}~\bibnamefont {Kruse}},
  \bibinfo {author} {\bibfnamefont {C.}~\bibnamefont {Peltz}}, \bibinfo
  {author} {\bibfnamefont {D.}~\bibnamefont {Engel}}, \bibinfo {author}
  {\bibfnamefont {A.}~\bibnamefont {Husakou}}, \bibinfo {author} {\bibfnamefont
  {T.}~\bibnamefont {Witting}}, \bibinfo {author} {\bibfnamefont
  {M.}~\bibnamefont {Ivanov}}, \bibinfo {author} {\bibfnamefont {M.~J.~J.}\
  \bibnamefont {Vrakking}}, \bibinfo {author} {\bibfnamefont {T.}~\bibnamefont
  {Fennel}},\ and\ \bibinfo {author} {\bibfnamefont {A.}~\bibnamefont
  {Mermillod-Blondin}},\ }\bibfield  {title} {\bibinfo {title} {{Origin of
  strong-field-induced low-order harmonic generation in amorphous quartz}},\
  }\href {https://doi.org/10.1038/s41567-020-0943-4} {\bibfield  {journal}
  {\bibinfo  {journal} {Nat. Phys.}\ }\textbf {\bibinfo {volume} {16}},\
  \bibinfo {pages} {1035} (\bibinfo {year} {2020})}\BibitemShut {NoStop}%
\bibitem [{\citenamefont {Wang}\ and\ \citenamefont {Du}(2021)}]{Wang2021f}%
  \BibitemOpen
  \bibfield  {author} {\bibinfo {author} {\bibfnamefont {G.}~\bibnamefont
  {Wang}}\ and\ \bibinfo {author} {\bibfnamefont {T.-Y.}\ \bibnamefont {Du}},\
  }\bibfield  {title} {\bibinfo {title} {{Quantum decoherence in high-order
  harmonic generation from solids}},\ }\href
  {https://doi.org/10.1103/PhysRevA.103.063109} {\bibfield  {journal} {\bibinfo
   {journal} {Phys. Rev. A}\ }\textbf {\bibinfo {volume} {103}},\ \bibinfo
  {pages} {063109} (\bibinfo {year} {2021})}\BibitemShut {NoStop}%
\bibitem [{\citenamefont {Vampa}\ \emph {et~al.}(2014)\citenamefont {Vampa},
  \citenamefont {McDonald}, \citenamefont {Orlando}, \citenamefont {Klug},
  \citenamefont {Corkum},\ and\ \citenamefont {Brabec}}]{Vampa2014}%
  \BibitemOpen
  \bibfield  {author} {\bibinfo {author} {\bibfnamefont {G.}~\bibnamefont
  {Vampa}}, \bibinfo {author} {\bibfnamefont {C.~R.}\ \bibnamefont {McDonald}},
  \bibinfo {author} {\bibfnamefont {G.}~\bibnamefont {Orlando}}, \bibinfo
  {author} {\bibfnamefont {D.~D.}\ \bibnamefont {Klug}}, \bibinfo {author}
  {\bibfnamefont {P.~B.}\ \bibnamefont {Corkum}},\ and\ \bibinfo {author}
  {\bibfnamefont {T.}~\bibnamefont {Brabec}},\ }\bibfield  {title} {\bibinfo
  {title} {{Theoretical Analysis of High-Harmonic Generation in Solids}},\
  }\href {https://doi.org/10.1103/PhysRevLett.113.073901} {\bibfield  {journal}
  {\bibinfo  {journal} {Phys. Rev. Lett.}\ }\textbf {\bibinfo {volume} {113}},\
  \bibinfo {pages} {073901} (\bibinfo {year} {2014})}\BibitemShut {NoStop}%
\end{thebibliography}%
\end{document}